\documentclass[12pt]{article}
\usepackage{geometry}                
\usepackage{graphicx}
\usepackage{amssymb,amsfonts, amsmath, amsthm, amstext}
\usepackage{subfigure}
\begin{document}

\title{Singularities, geodesics and Green functions\\
 in the BTZ black hole}
\author{Chen Yang\\
\text{}\\
\textit{\normalsize {Centre for Particle Theory}}\\
\textit{\normalsize{Department of Mathematical Sciences}}\\
\textit{\normalsize{University of Durham}}\\
\textit{\normalsize{Science Laboratories, South Rd,}}\\
\textit{ \normalsize{Durham, DH1 3LE, UK}}\\
\textit{\normalsize{email: chen.yang2@durham.ac.uk}}}
\maketitle
\begin{abstract}In the context of studying black hole singularities by the AdS/CFT correspondence, we study the BTZ black hole by a scalar field propagating on it and the corresponding two-point Green functions. We explore how  positions inside the horizon are encoded in the boundary theory. The main idea is to relate two different semi-classical approximations of the Green function and see how this indicates the bulk-boundary correspondence. From a key observation of Festucia and Liu, which is a frequency-geodesic identification,   we deduce a geodesic approximation from the saddle point approximation. As an application, we find saddles of the Green function and hence their corresponding geodesics. The conclusion is that some of these geodesics do go inside the horizon. This gives the possibility of resolving the singularity from the boundary theory.
 \end{abstract}

\newpage
\tableofcontents       
\section{Introduction}
Lack of satisfactory description of black hole singularities indicates the limitation of classical geometry. String theory as a generalization of general relativity is a candidate to resolve the singularities.  
One way to understand  the black hole singularities is using the so-called AdS/CFT correspondence conjectured by Maldacena \cite{maldacena-1999-38}. This conjecture  is motivated by a simple  comparison between pairs of decoupled theories occurred in some limit of superstring theory. One guiding principle of the correspondence is that the identification of isometry group of the bulk and the conformal group on the boundary. The proposal is the correspondence between an isometrically invariant theory on the bulk and a conformally invariant theory on the boundary. An extensive review on this subject is given in  \cite{aharony-2000-323}.

The motivation is simple, while the proof or even the exact statement of the conjecture  is still far from complete.  There are lots of efforts on matching the dual theories to  support the conjecture. Early examples are \cite{gubser-1998-428, witten-1998-2, banks-1998-, klebanov-1999-556}.  
 Instead of waiting for an ultimate proof, lots of us already enjoy the implication of the correspondence. The correspondence in principle allows one to study the bulk theory from the boundary point of view.  We will give a brief summary on the methods of studying bulk geometry  by the boundary CFT.

The original proposal and its support as \cite{klebanov-1999-556}  are both for the Euclidean signature of AdS space. Banks et. al. \cite {banks-1998-} and Balasubramanian et. al. \cite{balasubramanian-1999-59-2} independently give its generalization in the Lorenzian signature. We describe the correspondence in the example of a Klein-Gordon free scalar field on the bulk $AdS_{d+1}$. The (positive-frequency) modes solution is decomposed to two classes: normalizable modes and non-normalizable modes. On the bulk  Klein-Gordon quantum field theory, the non-normalizable modes are sources on the background and the normalizable modes are those which generate the Hilbert space of physical states in the particular background. Based on the principle that dual theories should have compatible symmetries required  by the identification between isometry group of the bulk and the conformal group on the boundary, the boundary operator related to a KG field on the bulk will be an operator coupled with a source on the boundary, which is from the restriction of  the non-normalizable part of the bulk field. 

At the quantum field theory level, the Hilbert space of the boundary theory is generated by the image of normalizable modes from the bulk. The annihilation and creation operators are the image of those from the QFT on the bulk. The corresponding vacua of the dual theories are understood as states that are annihilated by all annihilation operators in their  respective theories. 

The fundamental relation is the identification between partition functions of the dual theories. That is, the partition function of the KG theory on the bulk defined by  its classical action is equal to  the generating functional of boundary operator coupled with the source. This identification, in particular, implies the correspondence of two-point functions occurring in the dual theories.

After a general view of the correspondence for AdS space, we now give a brief summary on the correspondence for BTZ black hole \cite{banados-1993-48}, which is simply a quotient space of $AdS_3$ space. The  BTZ/CFT correspondence is studied at the level of modes solution of KG-field by Keski-Vakkuri \cite{keskivakkuri-1999-59} as a generalization of \cite{balasubramanian-1999-59-2}. 
 Maldacena in \cite{maldacena-2003-0304} lays out the BTZ/CFT correspondence in a Hartle-Hawking-Israel  state. Inspired by Israel's idea  in the subject of thermo-field dynamics of Schwarzschild black holes \cite{israel-1976}, Maldacena proposed that
 an AdS Schwarzschild like black hole can holographically be described by considering two identical, non-interacting copies of CFT in a particular entangled state. They are entangled in the sense that they communicate before their respective evolutions. The H-H prescription of gluing half of the Euclidean spacetime to half of the Lorenzian spacetime  defines the  Hartle-Hawking-Israel entangled state.  
 This H-H prescription  implicitly assumes the analytical continuation procedure, which is a defining feature of the original   Hartle-Hawking's Feynman propagator on a Schwarzschild black hole   \cite{hartle-1975}. Hence it is natural to regard it as a  Hartle-Hawking like state on a  AdS black hole. 
 Under this fixed initial state, one can talk about the correspondence between the dual quantum theories definitely.

Among various  objects in the dual theories, we  concentrate on the two-point correlation function. In the Euclidean signature AdS/CFT, Klebanov and Witten   \cite{klebanov-1999-556}  give the  sample calculation of obtaining the two-point function (in  the Poincar\'e vacuum) from the bulk and hence obtain the matching with the two-point function on the boundary as a support for the conjecture. The main step in 
 Witten's method is to give a bulk-boundary Green's function and so that for any boundary field, a unique bulk can be obtained. 
 In passing from Euclidean signature to Lorentzian signature, the new feature is appearance of normalizable modes \cite{balasubramanian-1999-59-2}. This breaks the uniqueness of finding the bulk field. However,
in the Hartle-Hawking-Israel state, the gluing process mentioned above implies the choice of  the bulk-to-boundary Green's function as  the analytic continuation from that in the Euclidean signature. Therefore, the whole procedure of finding the two-point function is just like the procedure in the Euclidean signature without ambiguity \cite{hemming-2002-0210, maldacena-2003-0304}. The  two-point correlation function in the Hartle-Hawking-Israel vacuum will be considered as a  Hartle-Hawking Green's function of BTZ black hole. 
This is the general framework of the correspondence of two-point functions. We will see how to use them to tell the geometry in a holographic manner.

The very first question on regard of studying bulk geometry from boundary point of view is  how to indicate the radial direction in the boundary theory. Earlier work in this direction are for example \cite{horowitz-1998-9804, maldacena-1998-9812} and  more recent attempts are as in
 \cite{balasubramanian-2000-61, louko-2000-62, hemming-2002-0210, maldacena-2003-0304, kraus-2003-67, fidkowski-2004-0402, kaplan-2004-, hamilton-2006-73}.  An important tool in this content is the geodesic approximation \cite{louko-2000-62} of the two-point Green function.   In the semi-classical limit, the two-point correlators can be represented in the form of Feynman paths integral of quantum mechanics, and the integration is dominated by certain geodesics. In this way, the geometric picture on the bulk appears. It is possible to probe singularities by studying the boundary objects which can be related to  geodesics  going inside the horizon. Works on this aspect are for example \cite{balasubramanian-2000-61, louko-2000-62,  fidkowski-2004-0402, kaplan-2004-}.

In the same stream, Festuccia and Liu \cite{festuccia-2006-0604} studied the $AdS_5$ black hole singularity. They
gave a method to match space-like geodesics in the bulk  with momentum space Wightman functions of CFT operators in the semi-classical limit. 
We study the two-point Green function of  a massive free scalar field in the Hartle-Hawking-Israel state on the BTZ geometry and its dual on the boundary of BTZ.  From a bulk point of view, we obtain  two-point function essentially by the method of modes sum. On one hand, the Green function,  written as the Fourier integration of the momentum space Green function, can be approximated by the methods of steepest descent \cite{Erdelyi}. The integration is dominated by its evaluation at saddles. On the other hand, if we write the Green function in the Feynman paths integral of quantum mechanics, it can be approximated by the geodesic approximation. The key to relate the two approximations is the observation of a frequency-geodesic relation. After a precise analytic continuation procedure, the relation becomes  mathematically well-defined. 

 This can be explained  in the following way. Consider this scalar field propagating on the BTZ black hole and its two-point function with the two points $A$ and $B$ inserted on distinct boundaries in the Penrose diagram. Taking semi-classical limit, it can be seen as a  wavepacket propagating along geodesics. This can also be explained as in the  Feynman path integration of quantum mechanics, where dominant terms are certain geodesics joining $A$ and $B$. It is plausible to make a term by term identification of the WKB approximation of the wave equation and the geodesic equation, both appearing as equations of motions in the semi-classical limit. After proper analytic continuation procedure, this identification gives a definite mapping between frequencies (saddles) and geodesics. Hence a geodesic approximation can be deduced from the properties of saddles. 
 
 Mathematically, this frequency-geodesic relation gives the following interesting property: 
Firstly, real frequencies parametrize Euclidean time separation of geodesics and these geodesics have real turning point outside the horizon. Secondly, purely imaginary frequencies parametrize Lorentzian time separation of geodesics and these geodesics have real turning point inside the horizon. The position space Green function in terms of Fourier transformation has an integration contour along real frequencies and saddles picked are near real axis, where the corresponding geodesics are with turning points outside the horizon
 
 Furthermore, if we analogously deform the integration contour from the real axis to the imaginary axis we get a new function, assuming that a regularization of the function exists. Saddles of the new function correspond to those geodesics with real turning point infinitely closed to the singularity. In this sense, we find signatures of the singularity on the bulk in the boundary correlation function. 
 
  We should make two remarks regarding the above method: First of all,  this is by no means the only way to make a geodesic approximation. For example different analytic procedure can lead to different dominating geodesics, while the correlation function remains the same. This kind of argument was made by Kraus, et. al. in \cite{kraus-2003-67}. They analytically continued the correlation function directly in two different procedures but found the same physical amplitude. 
Secondly, the frequency-geodesic relation does not have physical meaning beyond the semi-classical limit. The reason  is that the validity of this relation is supported by  the principle of wave-particle duality. This is only meaningful in the semi-classical limit. For example, for a frequency which is not at saddle point, the corresponding geodesic may not join the two  points in the two-point function.  

The plan for the rest of the paper is as follows. 
Section two will give a summary of the BTZ metric and the some remark on field quantization on BTZ space. In section three, we will give the BTZ/CFT correspondence and specify the choice of Hartle-Hawking-Israel state and  write down a Hartle-Hawking Green function by summation of  modes. Section four and five will give the explicit mapping between geodesics and saddles on the BTZ black hole at the  semi-classical limit by comparing the Hartle-Hawking Green function in two different approximations.  Section six is an application of this geodesic approximation.  Finally, we draw some conclusions. 

\section{The BTZ geometry and field quantization on BTZ}
\subsection{Classical construction of BTZ black hole}
We summarize the original construction of BTZ black hole of Banados, Henneaux, Teitelboim and  Zanelli \cite{banados-1993-48} and also give it in various coordinate systems. Extensive description of the geometry is also in the review \cite{aharony-2000-323}.
\subsubsection*{The $AdS_3$ space}
Since  the  BTZ black hole metric is obtained as a quotient metric from the $AdS_3$ metric. So we summarize the geometry of $AdS_3$ space first. 
The $AdS_3$ space is a hyperboloid in side the flat $\mathbb R^{2, 2}$, with metric 
$$ds_{\mathbb R^{2, 2}}^2=-dX_0^2-dX_{3}^2+ dX_1^2+dX_2^2, $$ 
defined by the equation, 
\begin{equation}
X_0^2+X_{3}^2- X_1^2-X_2^2=R^2.
\label{a1}
\end{equation}

The equation (\ref{a1}) can be solved by setting the following coordinate transformation from $X_i$'s to $(\tau, \rho, \Omega_1, \Omega_2)$,
$$X_0= R\cosh\rho \cos\tau,$$
$$X_{3}=R\cosh\rho \sin\tau, $$
$$X_1=R \sinh \rho\, \cos\psi,$$
$$X_2=R \sinh \rho\, \sin\psi,$$
where $0\leq\tau <2\pi$, $\rho\geq 0$ and $0\leq\psi<2\pi$ the solution covers the entire $AdS_{3}$ space once. Hence the coordinates $(\tau, \rho, \psi)$ are called the \textit{global coordinates} of the AdS.
The restriction of the metric $ds^2_{\mathbb R^{2, 2}}$ onto this hyperboloid is thus
\begin{equation}
ds_{AdS_{3}}^2= R^2(-\cosh^2\rho\,  d\tau^2+d\rho^2+\sinh^2 \rho\, d\psi^2).\label{ads}
\end{equation}
This is the $AdS_3$ space. 
This space has the topology of $S^1\times\mathbb R^{2}$ with $S^1$ the closed time direction in the $\tau$ direction. To avoid closed time like curve, we  may unwrap the circle to obtain a straight line $\mathbb R$.  This space is called the \textit{universal covering space of AdS}, dentoted as $CAdS_3$.

\subsubsection*{Conformal rescaling of $CAdS_3$ as half of ESU}
Since the causal structure has the property of  invariant under conformal rescaling of a metric, we can study the causal structure of $CAdS_3$ by studying that of a conformally rescaled metric, which will turn out to be the \textit{Einstein static universe}.  


We firstly  introduce the following coordinates.  We relate $\theta$ to $\rho$ by $\tan\theta=\sinh \rho,$ with $0\leq\theta<\pi/2$. Hence the metric (\ref{ads}) takes the form
\begin{equation}
 ds^2_{AdS_3}=\frac{R^2}{\cos^2\theta}(-d\tau^2+d\theta^2+\sin^2\theta d\psi^2),\label{a2}
\end{equation}
where  $-\infty<\tau<\infty$  and $0\leq\psi<2\pi$. 
This coordinate transformation is in particular a conformal transformation of the original CAdS. Now we conformally rescaling the metric to a new metric as
\begin{equation}
ds_{E/2}^2=-d\tau^2+d\theta^2+\sin^2\theta d\psi^2.
\label{a3}
\end{equation}
If $\theta$ takes value $0\leq\theta<\pi$ rather than $0\leq\theta<\pi/2$. This 
is a metric called the \textit{Einstein static universe (ESU)}, if the domain of $\theta$ is $[0, \pi]$.  Therefore we obtain a conformal rescaling of $AdS_3$ as half of an ESU.

\subsubsection*{Boundary of the conformal rescaled $CAdS_3$}
 We are interested in the boundary of the conformally rescaled $CAdS_3$ space. In the global coordinates of  the original $CAdS_3$, the boundary is at $\rho=\infty$. The corresponding boundary of the conformally rescaled (half of the Einstein static universe) would be at $\theta=\pi/2$ in the coordinate system $(\tau, \theta, \psi)$. In particular, the rescaled metric restricted on the boundary is give by
 \begin{equation}
 \label{einstein2}
 ds^2_{E/2}|_{\theta=\pi/2}=-d\tau^2+d\psi^2,\end{equation}
 where $-\infty<\tau<\infty, 0\leq\psi<2\pi.$ We see that the boundary is actually a whole ESU of dimension two. 

It is not hard to show that the boundary of the conformally rescaled $CAdS_3$ as in (\ref{einstein2}) is actually a conformal compactification of the Minkowski space $\mathbb R^{1, 1}$.
It is the identification of the isometry group of the conformally rescaled $CAdS_3$ and the conformal group of the conformally compactified Minkowski space that  plays fundamental role in the AdS/CFT correspondence.

\subsubsection*{The BTZ space}
The BTZ black hole metric will be obtained from the quotient metric of $CAdS_3$ by certain discrete isometry subgroup of the $SO(2, 2)$.
In general, if $\xi$ is a Killing vector on the spacetime, the transformation from any point $P$ to $\exp(s\xi) (P)$ in $CAdS_3$ is an isometry. The group $\{\exp(s\xi):s\in\mathbb R\}$ is an isometry group parametrized by $s$. If $s$ is restricted to $s=2\pi n$ for $n\in\mathbb Z$, then it parametrises a discrete subgroup. We  call it $\Xi(\xi)$. The quotient space $CAdS_3/\Xi(\xi)$ inherits a quotient metric from $CAdS_3$. Under the coordinates $(X_0, X^1, X^2, X_3)$, it is proposed that the Killing vector
$$\xi=r_+\left(X_0\frac{\partial}{\partial X_3}-X_3\frac{\partial}{\partial X_0}\right)$$ 
up to an isometric transformation. The BTZ  metric is defined to be the quotient metric induced from $CAdS_3$ through the discrete isometry subgroup $\Xi(\xi)$. However, to  exclude time-like curves on the quotient metric,  $\xi\cdot\xi>0$ needs to be satisfied everywhere on the spacetime. Hence the final BTZ black hole spacetime  is defined to be the BTZ metric on the region where $\xi\cdot\xi>0$. 
These regions can be divided into an infinite number of regions of two different types. 
\begin{enumerate}
\item
Region of type I,  $r_+^2<\xi\cdot\xi<\infty$ (the outer region).
\item
Region of type II, 
$0<\xi\cdot\xi<r_+^2$ (the inner region).
\end{enumerate}
We restrict to  two continuous regions of both types and introduce the coordinates $(\tilde t, r, \psi)$ as follows,
\begin{enumerate}
\item
Region I: $r_+<r,$
$$ X_0=\frac{r}{r_+}\cosh{r_+\psi},   \quad X_3=\frac{\sqrt{r^2-r_+^2}}{r_+}\sinh{r_+ \tilde t},$$
$$X^1=\frac{r}{r_+}\sinh{r_+\psi},\quad X^2=\frac{\sqrt{r^2-r_+^2}}{r_+}\cosh{r_+\tilde t}.$$
 \item
Region II: $r<r_+,$
$$ X_0=\frac{r}{r_+}\cosh{r_+\psi}, \quad X_3=-\frac{\sqrt{-r^2+r_+^2}}{r_+}\cosh{r_+ \tilde t},$$
$$  X^1=\frac{r}{r_+}\sinh{r_+\psi}, \quad X^2=-\frac{\sqrt{-r^2+r_+^2}}{r_+}\sinh{r_+\tilde  t}.$$
\end{enumerate}
In the coordinates $(\tilde t, r, \psi)$, the Killing vector is $\xi=\partial/\partial \psi$ and the discrete subgroup acts as taking the identification $\psi=\psi+2n\pi,$ for $ n\in\mathbb Z$. The BTZ black hole metric is thus the quotient metric by the isometry discrete subgroup generated by $\xi$  on the region where  $\xi\cdot\xi>0$,
\begin{equation}
\label{BTZ}
ds^2_{BTZ}=-(r^2-r_+^2)d\tilde t^2+\frac{1}{r^2-r_+^2} dr^2+r^2 d\psi^2
\end{equation}
with the identification $\psi=\psi+2n\pi,$ $ n\in\mathbb Z$. This is a black hole metric, in which the black hole singularity is actually a conic singularity, rather than a true curvature singularity. 

\subsubsection*{Causal structure of BTZ black hole}
To study the global structure of BTZ black hole, we introduce the Kruskal coordinates as follows,
In the region $0<r<r_+$, 
$$U=\sqrt{\frac{-r+r_+}{r+r_+}}\sinh\frac{r_+}{R^2}\tilde t;$$
$$V=\sqrt{\frac{-r+r_+}{r+r_+}}\cosh\frac{r_+}{R^2}\tilde t.$$
In the region $r_+<r<\infty$, 
$$U=\sqrt{\frac{-r+r_+}{r+r_+}}\cosh\frac{r_+}{R^2}\tilde t;$$
$$V= \sqrt{\frac{-r+r_+}{r+r_+}}\sinh\frac{r_+}{R^2}\tilde t.$$

In either region, we make the coordinate transformation from $(U, V)$ to $(p, q)$ as
$$U+V=\tan\frac{p+q}{2};$$
$$U-V=\tan\frac{p-q}{2}.$$

The result is that (i) $r=\infty$ is mapped to the lines $p=\pm\frac{\pi}{2}$, (ii) the singularity $r=0$ is mapped to the lines $q=\pm\frac{\pi}{2}$, and (iii) the horizon $r=r_+$ is mapped to $p=\pm q$. The $(U, V)$ coordinates give the Kruskal diagram and the $(p, q)$ coordinates give the Penrose digrams.

\subsection{QFT on BTZ}
\subsubsection*{ QFT in curved spacetime}
To define a quantum field theory in a Minkowski space time, one always has a distinct time direction to start with and this time direction together with the Poincar\'e symmetry of the spacetime is the key to define a unique quantum field theory. In particular, any two inertial observers in Minkowski spacetime agree on what is a vacuum state. 


The story on a general curved spacetime is different because it is possible to define infinity many inequivalent quantum field theories mathematically \cite{wald-1994}.
Take the example of quantization a  Klein-Gordon field again. If we pick a time direction $t$ as in the flat case, a quantum field theory can be defined as long as we find the set of positive frequency modes as eigen modes of the time Killing vector and the Klein-Gordon inner product on a Hilbert space spanned by these modes. Vacuum can also be defined by the annihilation operators appeared as coefficients of expansion of the field in the basis. However, we should say that different choice of $t$, the respective sets of positive modes may or may not span the same Hilbert space. In the later case, we obtain two distinct quantum field theories.

The lack of uniqueness of QFT should not be seen as a drawback curved spacetime, rather a sign of the nonzero curvature. The ambiguity arises only due to the incomplete (in the sense of unifying gravity and QFT) status of our theory. It is also such nonuniqueness  leads to the discovery of Hawking radiation of black hole \cite{hawking-1975}.

\subsubsection*{Hartle-Hawking state of a Schwarzschild black hole}
Despite the lack of  a general quantization procedure of fields on curved space, one can still by analyzing particular physical process and make choice of some physically meaningful QFT in a curved spacetime. One such example is the QFT in Hartle-Hawking state defined on a Schwarzschild black hole geometry.  It gives the physics of Hawking radiation of black holes. 

There are several equivalent explanation of the Hawking radiation.  We would like to take a historical routes.  Inspired by the nonuniqueness of vacuum states in curved spacetime, Hawking constructed a model so that black hole radiating particles is possible. For a  black hole with its apparent curvature singularity in the center, its asymptotic past and asymptotic future are still flat Minkowski space. Therefore,  a unique quantum field theory still can be defined on these regions. However, the choices of time direction in the region  connecting the two asymptotic areas are not equivalent, since they cannot be mapped to each other by a Poincar\'e transformation. Each choice of time direction will induce some ambiguity, which is caused by the curvature. These ambiguities are explained as some kind of quantum fluctuations of the curved spacetime. Add them up, Hawking proposes the probability of radiation of particles from a black hole \cite{hawking-1975}. 

Later Hartle and Hawking \cite{hartle-1975} by generating Feynman paths integral for quantum mechanics on flat space to that on  curved spacetime and obtain an alternative description of the black hole radiation.  The proof  is based on the construction of a Feynman propagator on the black hole obtained as the analytic continuation of its Wick rotated Euclidean Feynman path propagator.  
The observation  is that the rate of the black hole radiation obtained in this model is the same as the thermal radiation of a black hole located in a thermal bath. 

Inspired by Hartle and Hawking's observation, Gibbons and Perry \cite{gibbons-1976} provide another equivalent point of view to Hawking and Hartle's model. Seen from an observer standing in distance from the black hole singularity, the problem is treated  as a grand canonical ensemble of states of the field at a temperature. They in particular represent the Hartle-Hawking Feynman propagator in terms of a thermal Green function from the perspective of an outside observer. The thermal Green function is  obtained by mode sum at Hawking's temperature. 

Israel \cite{israel-1976}  provides a third alternative derivation of Hawking radiation based on the subject named thermo-field dynamics. In Israel's description, the two-point function can be generalized to the case when the two points are inserted on distinct boundaries  as proposed by Maldacena in \cite{maldacena-2003-0304} in the course of studying   AdS black hole in the AdS/CFT. The essential idea is that a  quantum field theory, with a physical Hilbert  space say $\mathcal H$ live on the right side of the Penrose diagram can be explained as a composite space $\mathcal H\otimes\tilde H$ where $\tilde H$ is defined to be a formal Hilbert space live on the left side of the Penrose diagram defined in such a way that for any operators from the right side, its expectation value in the composite vacuum is the same as its statistical averrage for an ensemble in thermal equilibrium. From first sight the dual Hilbert space $H$ is formal, however, it can be a full quantum field in its own right. 

\subsubsection*{Field quantization on AdS space}
Back to the field quantization on curved space time. Since the BTZ black hole is a quotient of AdS space, we want to comment on field quantization on AdS spaces \cite{avis-1978, breitenlohner-1982, mezincescu-1984, burgess-1985}.
There are two obstructions in quantization on AdS space. (1) it has closed time-like curve. This can be solved easily by considering CAdS space instead; (2) the spacial infinity of (C)AdS is time-like, which means that CAdS is not global hyperbolic and hence Cauchy's problem is not well-posed on this geometry. It is the time-like boundary that allows information leaking at infinity within finite time. This implies that the resulting quantum field theory is not a closed system and  Cauchy surface can not be well-defined. 

In the case when the field are coupled to the metric conformally, one can transform the problem of quantization of a field on ESU space, because the CAdS space can be conformally rescaled to half of an ESU space. (In the coordinate systems we use, the ``half'' means the range of $\theta$ is $0\leq\theta\leq \pi/2$ rather than $0\leq \theta\leq\pi$.) The quantization of field on CAdS can be obtained by the quantization of a field live in a box in ESU with the walls at $\theta=\pi/2$. The boundary condition added on the wall could be reflective boundary condition, which will give conservation of energy.
With this boundary condition added, the Cauchy problem  becomes well-defined and a unique quantization procedure can be carried out on CAdS. 
That is, the restricted positive-frequency modes which satisfy the particular boundary condition span a Hilbert space with the inner product as an integration over the Cauchy surface and  the quantization procedure is carried out  as normal.
For other possible boundary conditions so to give a well-defined quantization, one may refer to references above.

There are also various quantization procedures provided for AdS black holes \cite{lifschytz-1994, hyun-1994, ghoroku-1994, ichinose-1995-447, satoh-1997, natsuume-1998-13, klemm-1998-58, Hemming-2004}.
Depending on the system, boundary condition on the horizon or origin is needed, too.
In our case on the BTZ black hole, we use the boundary condition as the reflective boundary condition mentioned above. Because it is the normalizable condition in BTZ coordinates \cite{keskivakkuri-1999-59}, which is the natural one to impose in the content of BTZ/CFT correspondence. It is also the analogue outmodes as used in \cite{gibbons-1976} in the classical Hartle-Hawking radiation.  We are not attempting to give a quantization procedure of KG field on BTZ in this article. Rather, we specify the particular boundary conditions  and obtain a Green function at Hawking temperature by summation of  modes. This will be done explicitly in the next section. The resulting Green function  can be regarded as Hartle-Hawking Green's function on BTZ black hole. 
\label{two}

\section{BTZ/CFT correspondence and  Hartle-Hawking Green functions}
Assuming that we have a Hartle-Hawking Green function on the BTZ already and that the various equivalent explanations of Hartle-Hawking Green's function for Schwarzschild black can somehow be carried over to BTZ black hole, we want to obtain its various form interested us in the BTZ/CFT correspondence.

\subsection{The BTZ/CFT correspondence}
We summarize the BTZ/CFT correspondence proposed by \cite{maldacena-2003-0304, hemming-2002-0210}. Based on Israel's description  in the field of thermo-field dynamics on black holes \cite{israel-1976},  Maldacena  \cite{maldacena-2003-0304} proposes that the spacetime can be holographically described by  two identical, non-interacting copies of entangled CFTs on the boundaries.

Consider the  Euclidean  metric of the BTZ black hole with metric in the following form
$$ds^2_{EBTZ}=(r^2-1)d\tau^2+\frac{dr^2}{r^2-1}+r^2 d \phi^2,$$
where we have set $r_+=1$, with the identification $\phi=\phi+2\pi, $ and $ \tau=\tau+\beta,$
with $\beta$ the inverse of Hawking temperature of the black hole. 
Equivalently, if we write
$z=|z| e^{i\tau},$
then the metric take the form
$$ds^2_{EBTZ}=4\frac{dz d\overline{z}}{(1-|z|^2)^2}+\frac{(1+|z|^2)^2}{(1-|z|^2)^2}d\phi^2.$$

The boundary of the Euclidean black hole  is at $r=\infty$, the corresponding CFT is thus parametrized by 
 $(\tau, \phi)$. It is indeed a CFT at finite temperature  $\beta^{-1}$. 
In the $(z, \tau, \phi)$ coordinate, the boundary  is at $|z|=1$. 

The  Euclidean signature metric with $z=\tau+r$ can be
analytic continued the imaginary part of $z$ as $Im(z)=t_L$, where the imaginary part of $z$ containing the Lorentzian time, and now in the Lorenzian section of the whole complex $z$. If we  write
$z=-v;$ and 
$\overline z=u,$
we obtain the Lorentzian eternal black hole in the Kruskal coordinates $(u, v, \tilde\phi)$ as 
$$ds^2_{BTZ}=\frac{-4dudv}{(1+uv)^2}+\frac{(1-uv)^2}{(1+uv)^2}d\phi^2.$$
If we further define coordinates $t_L, x$ by $U=t_L+x$ and $V=t_L-x$, so the diagram in $(t_L, x, \phi=constant)$ is like the usual Kruskal diagram. 
In this coordinate the boundary $r=\infty$ in the AdS coordinates maps to the two hyperbola at $UV=-1$, the past and future singularity are at $UV=1$.


Back to the complex coordinate $z$, it contains both the Euclidean section and the Lorentzian section. Presumably, the two sections of spacetime are not related. Restricting on the Lorentzian signature, we have a particular state at time $t_L=0$. The two QFTs on the two sections will start to evolve with respect to their own Lorentzian time. A particular initial states at $t_L=0$ can be simply  given by the Euclidean section of the space time at $Im(z)=0$ (see Fig. 3 of  \cite{maldacena-2003-0304}). This is referred to as the \textit{HH prescription}. Thus defined  state is  called the \textit{Hartle-Hawking-Israel state}. We will always fix this initial state in our following analysis. 

Let us examine what the state is from the boundary point of view. Recall that  the Euclidean signature $AdS_3$ is a $3$-disk. Excluding the $S^1_{\phi}$, it is a $2$-disc,
 with the angular direction parametrized by the Euclidean time $\tau$. The boundary at $u=\infty$   is the boundary of the two-disc (3-disc if considering $\phi$ direction), which is the circle.   The Euclidean time restricted on constant $u$-slices always have the length  $\beta$. The HH-prescription says that the $Im(z)=0$ slice in the Euclidean black hole  is glued to the Lorentzian section  $I_{\beta/2}\times S^1_{\phi}$  at each constant $u$-slices. When restricted this identification  at the boundary, (in the Euclidean signature the boundary is $u=\infty$ half of the circle surrounding the two-disk; in the Lorentzian signature the boundary is simply the hyperbola $UV=-1$), we obtain two Lorenzian boundaries connected by half of a Euclidean circle.  The proposal  is that two CFTs live on two pieces of  the Lorenzian boundaries, with the initial states equivalent to a state live on half of the Euclidean circle $I_{\beta/2}\times S_{\phi}^1$.  These are two decoupled, entangled CFTs. They are entangled from their communication at the start. 
 
If we denote the two copies of the Hilbert space of the two CFT by $\mathcal H_1$ and $\mathcal H_2$, and the whole Hilbert space $\mathcal H=\mathcal H_1\times\mathcal H_2$ of the system. The initial wavefunction $|\Psi\rangle\in\mathcal H$ of the Lorentzian evolution is of the explicit form 
 \begin{equation}
 \label{HHstate}
 |\Psi\rangle=\frac{1}{\sqrt{Z(\beta)}}\sum_n \exp(-\beta E_n/2)|E_n\rangle_1\times|E_n\rangle_2,\end{equation}
 where the sum runs over all energy eigenstates and $Z(\beta)$ is the partition function of one copy of the CFTs at temperature $\beta^{-1}$. (\ref{HHstate}) is  the Harlte-Hawking-Israel state.

\subsection{Hartle-Hawking Green  functions}
\subsubsection*{Hartle-Hawking Green function in the bulk and the boundary}
With the specification of the particular vacuum states, we consider, in the semiclassical approximation, the correlation functions for insertions of operators 
  on distinct boundaries \cite{maldacena-2003-0304, hemming-2002-0210}. For the BTZ black hole the correlation functions of points on distinct boundaries are  decided by the analytic continuation, whose validity is guaranteed  by the particular choice of the Hartle-Hawking-Israel state. 
    
 Suppose a two-point Hartle-Hawking Green function with two points on the same boundary, say the  one on the right, is given as $ G((r, u_+, u_-), (r,  u'_+, u'_-))$  ($r\longrightarrow\infty$) where
 $u_+=\phi+t$ and $u_-=\phi-t.$
  The analytic continuation implies that when one point moves to the left boundary, the new Green function is obtained by substitute entities in $G$ by
 $$u_{\pm}\longrightarrow  u_{\pm}\mp\frac{i\beta}{2}.$$
 It is this analytic continuation implicitly commits one to consider a particular states of CFT defined on the disconnected BTZ boundaries and it is implied by  the  HH-prescription mentioned before. 
 
 Similar procedure can be carried out  from the boundary point of view, we will denote the Green function in the boundary theory by $\tilde G$. 
The simple relation between the bulk and boundary Hartle-Hawking Green functions is 
\begin{equation}
\label{adscft}
\tilde G((u_+, u_-), (u'_+, u'_-))=\lim_{z, z'\longrightarrow 0} z^{-2 h_+}z'^{-2h_+}G((z, u_+, u_-), (z', u'_+, u'_-)), \end{equation}
where $h_+=\frac{1\pm\sqrt{1+m^2}}{2}$ is the conformal dimension of the field restricted on the boundary. 
\label{three}

Following the discussion in  Section \ref{two},
we will give an explicit two-point Green function for the bulk by summation of modes. Due to the boundary condition of modes we choose, this can be understood as a Hartle-Hawking two-point function on the bulk.

\subsubsection*{Modes solution of the Klein-Gordon equation}
In the following, we will find the normalizable modes  mentioned above for a scalar field on the BTZ black hole.
The Klein-Gordon equation of a free scalar field $\Phi(r, t, \phi)$ propagating on the BTZ metric (\ref{BTZ}) is given as 
\begin{equation}
\label{KG}
\square\Phi=m^2\Phi,
\end{equation}
where $\square$ is the Laplacian operator defined by the metric.
We can solve this equation by the method of separation of variables. Write the ansatz of positive frequency modes as
$$f_{\omega, J}(t, r, \phi)= \exp(iJ\phi)\exp(-i\omega t) r^{-1/2} X_{\omega, J}(r),$$
where $J$ is integer.

The  radial part of the equation of motion is
\begin{equation}
\label{1}
-(r^2-1)^2\frac{d^2 X_{\omega, J}(r)}{dr^2}-2r(r^2-1)\frac{dX_{\omega, J}(r)}{dr}-\omega^2 X_{\omega, J}(r)+V(r) X_{\omega, J}(r)=0,
\end{equation}
where
$$V(r)=(r^2-1)\left(\frac{J^2+\frac{1}{4}}{r^2}+\nu^2-\frac{1}{4}\right),\qquad\nu^2=1+m^2.$$
Related to terms in Section \ref{three}, 
we have  $h_{\pm}=(1\pm\nu)/2$. 

A standard quantization procedure follows this would be (1) add sensible boundary condition on these modes at infinity and only pick a subspace of these positive frequency modes to construct a Hilbert space; (2)  specify  the Klein-Gordon inner product on the Hilbert space and further normalize the basis to obtain an orthonormal basis; (3) expand the field solution in this basis and promote the corresponding coefficients to annihilation and creation operators; (4) determining vacuum state by using annihilation operators;
(5) the corresponding Green function in this vacuum can be obtained by summation of these orthonormal modes. 
Unfortunately, it will be beyond the scope of this paper to finish these steps rigorously. To obtain the final Hartle-Hawking Green function we will only involve (1) and  (5). Such construction is by no means unique.

Equation (\ref{1}) can be found in terms of hypergeometric functions,
\begin{equation}
\label{RKG}
X_{\omega, J}(r)=r^{\frac{1}{2}}(1-r^2)^{-\frac{i\omega}{2}}\left(C_1 r^{-iJ}G(r)+C_2 r^{iJ}H(r)\right),
\end{equation}
where
$$
G(r)=F\left(\frac{-i\omega-iJ -\nu+1}{2}, \frac{-i\omega-iJ+\nu+1}{2}; 1-iJ, r^2\right),
$$
$$
H(r)=F\left(\frac{-i\omega+iJ-\nu+1}{2}, \frac{-i\omega+iJ+\nu+1}{2}; 1+iJ, r^2\right)
$$ and $C_1, C_2$ are constants.

We specify the boundary condition for the positive-frequency modes as follows, 

\begin{equation}
\label{bcs}
X(z)\longrightarrow
\begin{cases}
\exp(i\omega z)\exp(i\delta)+\exp(-i\omega z)\exp(-i\delta) & \text{ as } z\longrightarrow\infty\qquad(\text{horizon})\\
C(\omega,J)z^{\frac{1}{2}+\nu} & \text{ as } z\longrightarrow 0\qquad (\text{boundary})
\end{cases},
\end{equation}
where $\delta$ is some real constant and the coordinate $z$ is defined by
 $$z(r)=\frac{1}{2}\ln{(r+1)/(r-1)}.$$
 The boundary condition at infinity agrees with the reflective boundary condition in  \cite{hemming-2001-64} in global coordinates. It is also the same as the boundary condition satisfied by normalizable modes in \cite{keskivakkuri-1999-59}. Therefore, from the point of view of a well-defined QFT and an AdS/CFT related modes, this boundary condition is natural. Furthermore, these modes when   restricted to $U=0$ ($V=0)$ have an expansion in terms of $e^{-i\omega V}( e^{-i\omega U})$ with $\omega>0$ and hence are related to Kruskal modes in the same way as out modes related to Kruskal modes, which is the defining modes of the Hartle-Hawking state of a Schwarzschild black hole. With all these inspections, we may regard the summation of these modes at Hawking temperature as a Hartle-Hawking like Green function of the BTZ black hole. The boundary condition we imposed at the horizon seems less transparent, except technically easy, and needs  a justification of its physical meaning.

With these boundary conditions, we are able to fix constants $C_1$ and $C_2$. The detailed calculation is in the appendix. The coefficient
\begin{eqnarray}
\label{cw}
C(\omega, J)^2&=&\frac{\exp(\pi\omega)(\exp(2\pi\omega)-1)\omega}{2\pi \Gamma(1+\nu)^2}\Gamma\left(\frac{1}{2}(i\omega+iJ+\nu+1)\right)\nonumber\\
&&\Gamma\left(\frac{1}{2}(-i\omega+iJ+\nu+1)\right)\Gamma\left(\frac{1}{2}(i\omega-iJ+\nu+1)\right)\nonumber\\
&&\Gamma\left(\frac{1}{2}(-i\omega-iJ+\nu+1)\right)
\end{eqnarray}
 at boundary\footnote{ This constant is worked out with the help of Mukund Rangamani.}, which will be used.

\subsubsection*{Hartle-Hawking Green function by modes sum}
The Wightman  function at zero temperature is given by the modes sum,
\begin{eqnarray*}
G^+((t, r, \phi), (t', r', \phi'))
&=&\int_{-\infty}^{\infty} d\omega\sum_{ J}\exp(i J(\phi'-\phi)-i\omega(t'-t))r^{-1/2}r'^{-1/2}\\
&&X_{\omega, J}(r)X_{\omega, J}(r').
\end{eqnarray*}
Note that $\omega$ is not quantized under the boundary condition (\ref{bcs}). 
On the other hand, the Fourier integration representation of $G^+((t, r, \phi), ( t', r', \phi'))$ is
\begin{eqnarray*}
G^+((t, r, \phi), (t', r', \phi'))&=&\int_{-\infty}^{\infty}d\omega\sum_J\exp(i J(\phi'-\phi)-i\omega(t'-t))\mathcal G^+(\omega, J; r, r').
\end{eqnarray*}
Comparing coefficients in the two integrations above, the Wightman function in the momentum space is
\begin{equation}
\label{momentum}
\mathcal G^+(\omega, J; r, r')=r^{-1/2}r'^{-1/2} X_{\omega, J}(r)X_{\omega, J}(r').
\end{equation}
The Wightman function $\mathcal G^+_{\beta}(\omega, J; r, r')$ at the Hawking temperature $T_H=\frac{1}{\beta}$, where $\beta=2\pi$,
 can be obtained from the Wightman function at the zero temperature by \cite{birrell},
$$
\mathcal G^+_{\beta}(\omega, J; r, r')=\frac{\mathcal G^+(\omega, J; r, r')}{1-\exp(-\beta\omega)}=\frac{\exp(\beta\omega)}{\exp(\beta\omega)-1}r^{-1/2}r'^{-1/2}X_{\omega, J}(r)X_{\omega, J}(r').
$$
This is the momentum space Hartle-Hawking Green function. 
We can pass the relation between position space two-point functions  (\ref{adscft}) to the momentum space two-point functions.
 The boundary Green function can be obtained as,
\begin{eqnarray*}
\tilde{\mathcal G}_{\beta}^+(\omega, J)&=&\lim_{r, r'\longrightarrow \infty} r^{2h_+}  r'{}^{2h_+}\mathcal G_{\beta}^+(\omega, J; r, r')\nonumber\\
&=&\lim_{r, r'\longrightarrow\infty} r^{1+\nu}r'^{1+\nu}\frac{\exp(\beta\omega)}{\exp(\beta\omega)-1}r^{-1/2}r'^{-1/2}C(\omega, J)^2 z^{1/2+\nu}z'^{1/2+\nu}\nonumber\\
&=&\frac{\exp(\beta\omega)}{\exp(\beta\omega)-1} C(\omega, J)^2.
\end{eqnarray*}
Substituting (\ref{cw}) in, we obtained the boundary thermal Green function in the momentum space,
\begin{eqnarray}
\label{3}
\tilde{\mathcal G}_{\beta}^+(\omega, J)&=&
\frac{\exp(3\pi\omega)\omega}{2\pi\Gamma(1+\nu)^2}\Gamma\left(\frac{1}{2}(i\omega+iJ+\nu+1)\right)\Gamma\left(\frac{1}{2}(-i\omega+iJ+\nu+1)\right)\nonumber\\
&&\Gamma\left(\frac{1}{2}(i\omega-iJ+\nu+1)\right)\Gamma\left(\frac{1}{2}(-i\omega-iJ+\nu+1)\right).
\end{eqnarray}

\begin{figure}[hpt]
\centering
\includegraphics[width=2.5in]{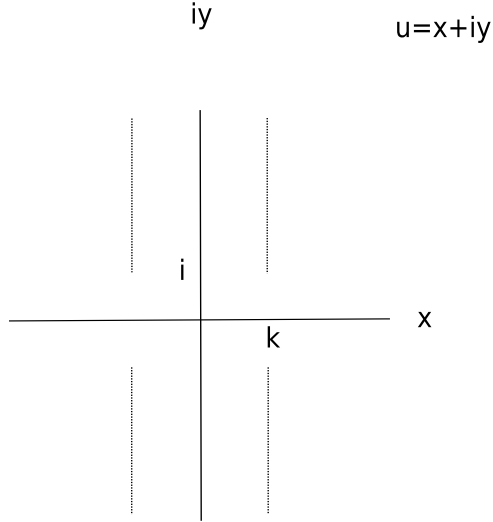}
\caption[]{Poles of $\tilde{\mathcal G}_{\beta}^+(u, k)$ in the large $\nu$ limit.}\label{fit16}
\end{figure}

To analytically continue this function by extending the domain of $\omega$ from the real line to the complex $\omega$-plane, we need to identify singular points of $\tilde{\mathcal G}_{\beta}^+(\omega, J)$. They are located at
\begin{equation}
\label{3-1}
\{\omega=\pm J\pm(1-\nu)i\mp2Ni\}\cup\{\omega=\pm J\mp(1-\nu)i\pm 2Ni\},\end{equation}
where $N$ is any positive integer. For later use, we change the scale of variables as
$u=\omega/\nu$ and $k=J/\nu$ and further assume that $k$ is a positive real number. This assumption will be made clear later by relating it to the assumption of $L=iL_I$ where $L_I>0$ in Section \ref{2.2}. Singular points  (\ref{3-1}) in the large $\nu$ limit are located at
\begin{equation}
\{u=\pm k\pm i\pm 2ni\}\cup\{u=\pm k\mp i\mp2ni\},
\label{cuts}
\end{equation}
where $n=N/\nu$ and  $N$ is any positive integer.

On the complex $u$-plane, the distance between poles become very small as $\nu\longrightarrow \infty$ and the four lines of poles become branch cuts. This is illustrated in Fig. \ref{fit16}. These are the branch cuts for analytical continuation of $\tilde{\mathcal G}_{\beta}^+(u, k)$.

\section{Geodesic approximation and Saddle point approximation of the Green function I}
We want to see how the exact boundary correlation function probe the bulk geometry. The main idea is to relate correlation functions to geodesics on the bulk. To achieve this, we consider two different semi-classical approximations of the two-point Green function.
The first method is the geodesic approximation. Indeed, wavepackets propagate like particles along geodesics in the semi-classical limit. 

\begin{figure}[hpt]
\centering
\subfigure[]
{
\includegraphics[width=2.8in]{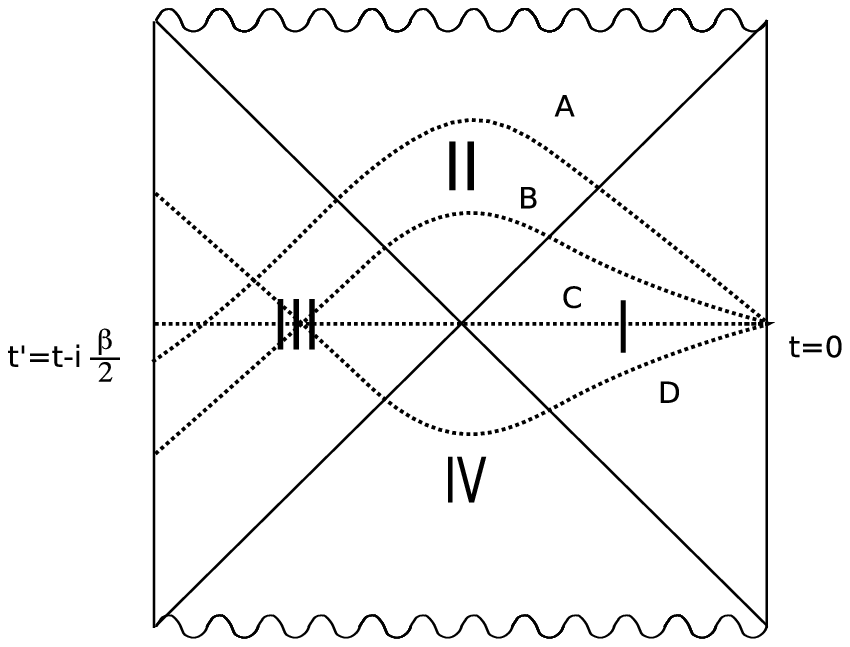}
}
\hspace{1cm}
\subfigure[]
{
\includegraphics[width=2in]{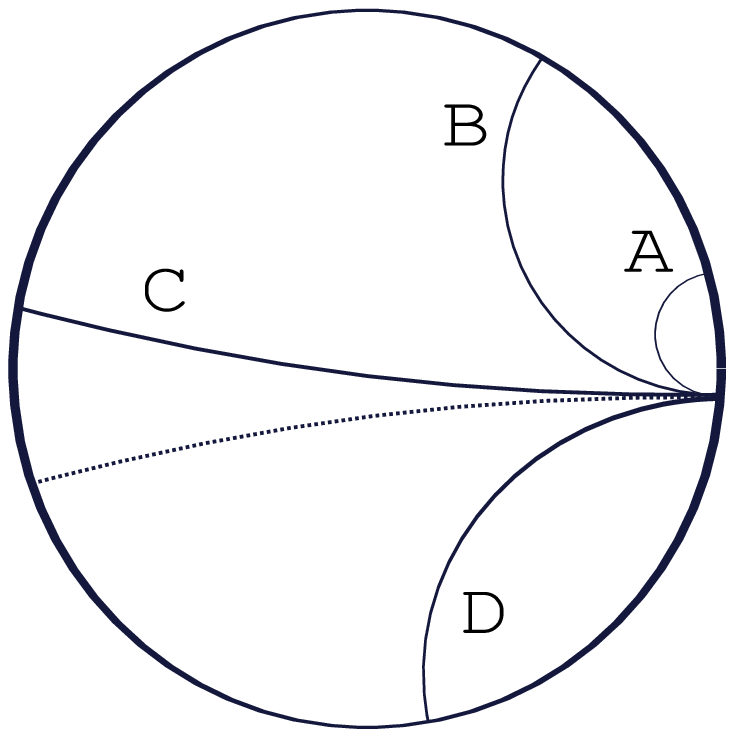}
}
\hspace{3cm}
\caption[]{ (a) Space-like geodesics joining points on distinct boundaries with time separation $t'-t-i\beta/2$ in the Lorentzian section of the spacetime.  Lines of equal Lorentzian time are
straight lines through the centre, the time being given by the slope. 
 Lines of equal radius $r$ are hyperbolas in each region (not shown), with asymptotics at the cross in the middle. The upper and lower waved lines are the future and past singularities. The upper part of the cross is the future horizon $r=1$ and the lower part of the cross is the past horizon. The left and right straight lines are two distinct boundaries at $r=\infty$. Each wedge is associated with a constant Euclidean time. Region I is of time $0$, region II is of time $-i\beta/4$, region III is of time $-i\beta/2$ and region IV is of time $-3i\beta/4$. The jumping  of $-i\beta/4$ comes from the fact that integration of proper time from one region to the next picks up a pole at the horizon with a  factor of $-i\beta/4$. $A, B, C$ and $D$ all have Euclidean time separation $-\beta/2$. As $u$ varies from $-i\infty$ to $0$ along the imaginary axis, geodesic moves from $A$ to  $B$ and then to $C$. Geodesic $D$ corresponds to $u_I>0$. (b) Space-like geodesics in the Euclidean section.  The radial direction of the spacetime is along the radial direction of the circle. The centre of the circle is chosen to be at the horizon, $r=1$ and the boundary of circle is where $r=\infty$. The Euclidean time $\tau$ is in the angular direction with period $2\pi$. As $u$ varies from $+\infty$ to $0^+$, the corresponding geodesic changes from $A$ to $B$, and then to $C$. Note that $C$ turns back before it reaches the horizon at the centre. Geodesic $D$ stands for a geodesic with $u<0$.
}
\label{fit11}
\end{figure}
With respect to the decomposition of complex time separation into the Lorentzian and Euclidean sections, i.e. $t+i\tau$, the BTZ spacetime can   be projected onto the two sections as shown in Fig. \ref{fit11}.

We are interested in the case when the two points are inserted on distinct boundaries of the Penrose diagram, describing the Lorentzian section of the spacetime. Geodesics joining distinct boundaries are like $A$, $B$, $C$  and $D$ in Fig. \ref{fit11} (a).

The second approach is by the steepest descent method to evaluate the correlation function as Fourier integration of the momentum space Green function. For points inserted on distinct boundaries in the Lorentzian section space-time, the boundary two-point function can be obtained from the Wightman functions by changing the time variable from $t'$ to $t'-i\beta/2$. That is,
\begin{equation}
\label{g12}
\tilde G_{12}(t, t'; k)=\int_{-\infty}^{\infty}\exp\left(-i\nu u\left(t'-t-i\frac{\beta}{2}\right)\right)\tilde {\mathcal G}_{\beta}{}^{+}(u, k).
\end{equation}
The asymptotic expansion of the integration is dominated by saddle points. The goal is to relate the saddles and the geodesics, both appeared as dominating objects in the semi-classical limit, so that we can deduce a geodesic approximation from the saddle point approximation. In other words, we want to make the following geodesic approximation explicitly defined in the semi-classical limit,
\begin{equation}
\label{cor3}
\tilde G_{12}(t, t'; k)=\sum_i \exp(-m (\textsf{l}_i+f_i(k))),
\end{equation}
where $i$ indicates different dominant geodesics, $m$ is the mass of the scalar field and $\textsf{l}_i$ is proper length of the $i$-th dominant geodesic joining the two points and $f_i(k)$ is a function indicating the contribution from the angular momentum $k$. We will firstly give the notion of geodesics on the BTZ black hole, secondly sketch the method of steepest descent and thirdly provide a formal proof of (\ref{cor3}). We will deduce an explicit geodesic approximation in the next section.
\subsection{Some formula related to geodesics on the BTZ black hole}
On the BTZ metric (\ref{BTZ}), there are two particular  Killing vectors, $\partial_t=(1, 0, 0)$ and $\partial_{\phi}=(0, 0, 1)$.
Their corresponding conserved quantities are energy $E^2$ and angular momentum $L^2$. Since energy and angular momentum are conserved on geodesics, we can write down geodesic equations in terms of $E$ and $L$.  Let $\vec x(\lambda)=(t(\lambda), r(\lambda), \phi(\lambda))$ be a space-like geodesic with the affine parameter $\lambda$.  Its geodesic equation can be written as
\begin{equation}
\label{geo2}
\frac{1}{r^2-1}\left(\frac{dr}{d\lambda}\right)^2+\frac{L^2}{r^2}-\frac{E^2}{r^2-1}=1,
\end{equation}
where 
$E=(r^2-1)dt/d\lambda$ and $L=r^2d\phi/d\lambda.$
\begin{figure}[hpt]
\centering
\includegraphics[width=3in]{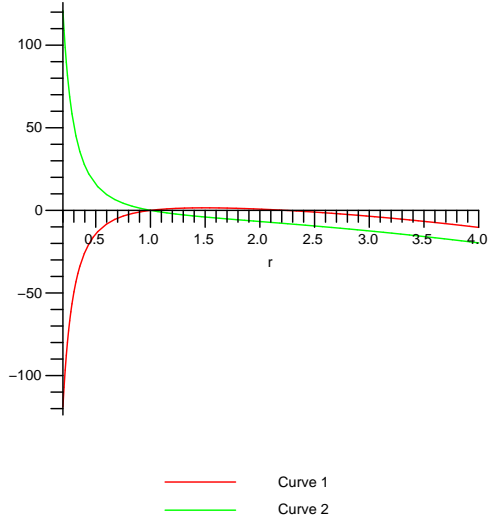}
\caption[]{Potential v.s. radius with  different $L$.}\label{fit21}
\end{figure}
If we define an effective potential by 
\begin{equation}
\label{potential}
V_{eff}=-(r^2-1)\left(1-\frac{L^2}{r^2}\right),
\end{equation}
then (\ref{geo2}) gives
$(dr/d\lambda)^2=E^2-V_{eff}.$ This can be thought of as the equation of motion of a particle of energy $E^2$ in a potential $V_{eff}$. 
Observe that the potential (\ref{potential}) behaves differently according to the different values of $L$. 

See Fig. \ref{fit21}. 
Imagine a particle of energy $E^2$ coming from boundary towards the singularity.
When $L$ is a real number, the potential is represented by the Curve 1 of Fig. \ref{fit21}. We can see that with big enough $E^2$, the particle will go inside the horizon and be trapped there. However, when $L$ is a purely imaginary number, the potential is represented by the Curve 2. We can see that no matter how big $E^2$ is, the particle has to bounce back. When $E^2$ is infinitely large, the turning point of the geodesic becomes infinitely closed to the singularity. In this case, the returned particle may carry signatures of the singularity. Therefore, we want to consider the case when $L$ is a purely imaginary number. For simplicity, we assume $L=iL_I$ for $L_I>0$. \label{2.2}
The turning point $r_c$ of  such particles can be solved as solutions of the equation
\begin{equation}
\label{turngeo}
E^2+r^2-L^2-1+\frac{L^2}{r^2}=0.
\end{equation}
The proper time \textsf{t}, proper length \textsf{l} and proper angular displacement \textsf{d} of a space-like geodesic with energy $E^2$ and angular momentum $L^2$ are as follows,
\begin{equation}
\label{geo3-1}
\textsf{t}(E, L)=2\int_{r_c}^{\infty}\frac{E}{(r^2-1)\sqrt{E^2+(r^2-1)\left(1-\frac{L^2}{r^2}\right)}}dr
\end{equation}
\begin{equation}
\label{geo3-2}
\textsf{l}(E, L)=2\int_{r_c}^{\infty}\frac{1}{\sqrt{E^2+(r^2-1)\left(1-\frac{L^2}{r^2}\right)}}dr
\end{equation}
\begin{equation}
\label{geo3-3}
\textsf{d}(E, L)=2\int_{r_c}^{\infty}\frac{L}{r^2\sqrt{E^2+(r^2-1)\left(1-\frac{L^2}{r^2}\right)}}dr,
\end{equation}
where the integrations are all over the contour along the real $r$ axis from $r_c$ to positive infinity.

\subsection{Steepest descent methods}
We will briefly summarize the steepest descent methods from \cite{Erdelyi, Bleistein}. It is a technique to get asymptotic expansions at the large $\nu$ for integrations in the form,
\begin{equation}
\label{form}
I(\nu)=\int_C g(u)\exp(\nu h(u))du,
\end{equation}
where $C$ is a contour on the complex $u$-plane, and $g(u)$ and $h(u)$ are holomorphic functions. The points of $u$-plane where $dh(u)/du=0$ are called \textit{saddle points}. They are saddle points on the real surface representing $|\exp(\nu h(u))|$, which is seen as a function of the two real variables $x$ and $y$ with $x=Re(u)$, $y=Im(u)$.  Curves along which $Im(\nu(h(u))$ is constant are called \textit{steepest descent paths}. Along such curves,  $|\exp(\nu h(u))|$ changes as rapidly as possible. In other words, they are the gradient lines of $|\exp(\nu h(u))|$. The saddle point is a stationary point of the function $|\exp(\nu h(u))|$ along the steepest descent path. The method of steepest descents consists in deforming the path of integration $C$ so as to make it coincide as far as possible with arcs of steepest paths. In this way, the integration is transformed to a real integration of the Laplacian type and the Laplace method may be used to evaluate the integral asymptotically. 
We summarize the methods in the following steps:
\begin{enumerate}
\item
Identify the saddle points $u_0$ as zeros of  $dh(u)/du=0$.
\item
Determine degree $n$ of saddle point $u_0$ such that
$$\frac{d^q h(u)}{du^q}|_{u_0}=0, \qquad q=1, 2, \dots, n-1, \qquad \frac{d^n h(u)}{du^n}|_{u=u_0}=a \exp(i\alpha), \qquad a>0$$
for some real number  $\alpha$.
\item
Determine directions of steepest descent of each saddle point:
If $u-u_0=\rho \exp(i\theta)$, then  steepest descent directions are
$$\theta=-\frac{\alpha}{n}+\frac{(2 p+1) \pi}{n}, \qquad p=0, 1, \dots, n-1,$$
and  steepest asscent directions are
$$\theta=-\frac{\alpha}{n}+\frac{2 p \pi}{n}, \qquad p=0, 1, \dots, n-1.$$

\item
Justify, via Cauchy's integral theorem, the deformation of the original contour of integration $C$ onto one or more of the paths of steepest descent. 
\item
Determine the asymptotic expansions of the deformed integrals through Laplace methods.
\end{enumerate}
For our interests of finding dominating saddles and eventually their corresponding geodesics, we do not need to go through the fifth step. 
The rest of the steps listed will become clearer when we consider our particular examples.

\subsection{Saddle points and the geodesic approximation}
We will formally deduce (\ref{cor3}) for now and provide a mathematically explicit description in the next subsection.
Consider the Fourier integration of the two-point function (\ref{g12}) in the large $\nu$ limit, 
\begin{equation}
\label{twopt}
\tilde G_{12}(t, t'; k)=\int_{-\infty}^{\infty} du \exp\left(-iu\nu \left(\Delta t-i\frac{\beta}{2}\right)\right)\exp(2\nu Z(u, k)),
\end{equation}
where $\Delta t=t'-t$ and we write $\tilde{\mathcal G}_{\beta}^{+}(u, k)=\exp(2\nu Z(u, k))$ for some $Z(u, k)$ so that the Fourier integration is transformed into the form of (\ref{form}).  We can analytically continue the function
 $\tilde{\mathcal G}_{\beta}^{+}(u, k)$ from the real $u$-axis to the  complex $u$-plane without difficulty by branch cutting  pole lines illustrated in Fig. \ref{fit16}. Complex saddles can thus be obtained by solving the saddle point equation. The evaluation of the integrand at the dominating saddles is dominant in the asymptotic expansion of \ref{g12}. On the other hand, we ca use the language of Feynman paths integration of quantum mechanics for (\ref{twopt}) in the semi-classical limit.  There will be certain geodesics joining the two points to dominate the integration in the spirit of wave-particle duality. Since the saddles and geodesics appear respectively as dominant terms in two different semi-classical approximations of the same correlation functions, we want to make a link between them. The first  attempt is to simply make a term by term identification, by comparing units, between the wave equation and the geodesic equation  \cite{festuccia-2006-0604}. 

To obtain the wave equation in the semi-classical limit, we use the ansatz $X(z)=\exp(\nu S(z))$ for the radial part of the Klein-Gordon field. Equation (\ref{1}) can thus be written in terms of $S(z)$. We substitute in the expansion  $S(z)=S_0(z)+1/\nu S_1(z)+1/\nu^2 S_2(z)+\cdots$, in which 
the $S_0(z)$ will be the leading term in the large $\nu$ limit.  The radial part of the  Klein-Gordon equation can further be written in terms of $S_0(z)$,
\begin{equation}
\label{4}
\frac{1}{r^2-1}\left(\frac {dS_0(z)}{dz}\right)^2-\frac{k^2}{r^2}+\frac{1}{r^2-1}u^2=1.
\end{equation}
On the other hand, recall the geodesic equation (\ref{geo2}), describing space-like geodesics  joining points on the boundary of energy $E^2$ and angular momentum $L^2$. 
The assumption of equivalence between equations (\ref{4}) and (\ref{geo2}) suggests the following identification:
\begin{equation}
\label{ast}
\frac{d S_0}{dz}\longleftrightarrow \frac{dr}{d\lambda}, \qquad
k\longleftrightarrow i L, \qquad
u\longleftrightarrow i E.
\end{equation}
Observe that the relation between $L$ and $k$ is consistent with the assumptions that $L=iL_I$ for $L_I>0$ and that $k$ is positive.

Now we will show that (\ref{ast}) implies the formalism of the geodesic approximation of propagators as in (\ref{cor3}). Write the momentum space Green function in terms of $S_0(z)$, i.e., 
\begin{equation}
\label{semi}
\tilde{\mathcal G}_{\beta}^+(u, k)
=\lim_{r, r'\longrightarrow\infty} r^{1+\nu}r'^{1+\nu}\frac{\exp(\beta\omega)}{\exp(\beta\omega)-1}r^{-1/2}r'^{-1/2} \exp(\nu S_0(z))\exp(\nu S_0(z'))+\cdots
\end{equation}
and define $Z_0(u, k)=\lim_{z(r)\longrightarrow 0} S_0(z)$. The semi-classical approximation of momentum space Green function is thus
$\tilde{\mathcal G}^+_{\beta}(u, k)\sim \exp(2\nu Z_0(u, k))$ where $\nu>>1$.
We apply the identification (\ref{ast}) and change coordinates from $z$ to $r$ in the integration. We obtain
\begin{eqnarray}
\label{6}
Z_0(u, k)&\stackrel{\cdot}{=}&\lim_{r\longrightarrow\infty}\int_{z(r_c)}^{z(r)}\frac{dS_0}{dz}dz\nonumber\\
&\stackrel{\cdot}{=}&\lim_{r\longrightarrow\infty}\int_{z(r_c)}^{z(r)}\sqrt{(r^2-1)\left(1+\frac{k^2}{r^2}-u^2\right)}dz\nonumber\\
&\stackrel{\cdot}{=}&iu\cdot\frac{1}{2}\textsf{t} (u, k)- \frac{1}{2}\textsf{l} (u, k) +\frac{1}{2}L  \textsf{d} (u, k),
\end{eqnarray}
where $r_c$ is the solution of 
\begin{equation}
\label{turningpt}
-\frac{k^2}{r^2}+\frac{1}{r^2-1}u^2=1
\end{equation}
and $\textsf{t}$, $\textsf{l}$, $\textsf{d}$ are from (\ref{geo3-1}),  (\ref{geo3-2}) and (\ref{geo3-3}), respectively. Also notice that equation (\ref{turningpt}) agrees with equation (\ref{turngeo})  under the assumption (\ref{ast}). Here ``$\stackrel{\cdot}{=}$'' reminds us that these are formal equalities. We will fix an analytic continuation procedure to make these equalities exact later. We make some immediate remark regarding (\ref{6}). It implies that
$$\frac{2\partial Z_0(u, k)}{\partial (-iu)}=-\textsf{t}(u, k), \qquad \frac{2\partial Z_0(u, k)}{\partial (-ik)}=\textsf{d}(u, k).$$
If we regard the proper length $\textsf{l}(\textsf{t}, \textsf{d})$ as a function of $(\textsf{t}, \textsf{d})$ instead of $(u, k)$, then the above identities imply that 
$2Z_0(u, k)$ and $\textsf{l}(\textsf{t}, \textsf{d})$ are related by a Legendre transformation.
Geometrically, a Legendre transformation maps the graph of a function to the family of tangents to the graph. In our case, if the graph is $\textsf{l}(\textsf{t}, \textsf{d})$ then the tangents are $d\textsf{l}(\textsf{t}, \textsf{d})/d\textsf{t}=iu$. Alternatively, if the graph is $2Z_0(u, k)$ then the tangents are $2dZ_0(u, k)/du=-i\textsf{t}$. 
As Lagrangian and Hamiltonian we have the mathematical relation
\begin{equation}
\label{legendre}
2Z_0(u, k)+\textsf{l}(\textsf{t}, \textsf{d})=iu\textsf{t}-ik\textsf{d}
\end{equation}
and hence
\begin{equation}
\label{zuk}
\frac{d2 Z_0(u, k)}{du}=i\textsf{t}.
\end{equation}
We further consider the decomposition of the identification $-iu=(r^2-1)d\textsf{t}/d\lambda$ into the real and imaginary parts, and take integration respectively. We obtain
$$\textsf{t}_E=\int\frac{ -u_R}{r^2-1}d\lambda+C_E, \qquad \textsf{t}_L=\int \frac{u_I}{r^2-1} d\lambda+C_L,$$
where $C_E$ and $C_L$ are constants of integration. We will fix them by fixing a reference geodesic $G_0$. We choose it as the geodesic with time separation $-i\beta/2$, which is the same as geodesic $C$ in Fig. \ref{fit11} (a). So knowing $u_R$ and $u_I$ is equivalent to knowing $\textsf{t}_E$ and $\textsf{t}_L$.
In other words, we assign $u=0$ to the geodesic $G_0$ and use other $u$ to measure how other geodesics are different from $G_0$ in proper time separation.  For example, if $u_I=0$, then the change of the proper time from that of $G_0$ is zero in the Lorentzian section. If $u_R$ is equal to zero, then the change of the proper time from that of $G_0$ is zero in the Euclidean section, that is to say the Euclidean time separation will always be the constant $-i\beta/2$.

Now come back to the semi-classical approximation of  $\tilde{\mathcal G}_{\beta
}^{+}(u, k)$ in (\ref{semi}). By applying (\ref{6}), we get
\begin{eqnarray*}
\tilde{\mathcal G}^+_{\beta}(u, k)&\sim&\exp(\nu(iu\cdot\textsf{t} (u, k)- \textsf{l} (u, k) +L  \textsf{d} (u, k))).
\end{eqnarray*}
The position space boundary two-point Green function (\ref{g12}) is therefore,
\begin{eqnarray}
\label{green1}
\tilde G_{12}(t, t'; k)
&\sim&\int_{-\infty}^{\infty} du \exp\left(-i  u\left(t'-t-i\frac{\beta}{2}\right)\nu\right)\nonumber\\
&&
\exp\left(\nu\left(iu\cdot\textsf{t} (u, k)- \textsf{l} (u, k) +L  \textsf{d} (u, k)\right)\right).
\end{eqnarray}
Evaluating by steepest descent methods, the integration (\ref{green1}) will be dominated by its evaluation at saddle points, which are solutions of
$2dZ_0(u, k)/du=i(\Delta t-i\beta/2).$
By identity (\ref{zuk}),  the saddle point equation implies
\begin{equation}
\label{timesep}
\textsf{t}=\Delta t-i\frac{\beta}{2}.
\end{equation}
That is to say, saddle points as frequencies under the identification (\ref{ast}) will be identified as geodesics with proper time separation equals to $\Delta t-i\beta/2$. This agrees with the physical picture we have in mind. However, under the same identification,   frequencies which are not saddles correspond to geodesics with time separation not necessarily $\Delta t-i\beta/2$.  This means that those geodesics may not join the two points which appeared in the two-point function.  The reason is that the relation (\ref{ast}) is obtained in the semi-classical limit when the wave-particle duality holds. We have no reason to  expect it to be physically meaningful beyond this limit. Finally, by substituting (\ref{legendre}) into (\ref{green1}) and  applying (\ref{timesep}), we have
\begin{eqnarray*}
\tilde G_{12}(t, t'; k)&\sim&\int_{-\infty}^{\infty} du \exp\left(-i  u\left(\Delta t-i\frac{\beta}{2}\right)\nu\right)\exp(\nu(-\textsf{l}(\textsf{t}, \textsf{d})+iu\textsf{t}-ik\textsf{d}))\nonumber\\
&\sim&\sum_{u_i}\exp(-m(\textsf{l}_i(\textsf{t}_i, \textsf{d}_i)+ik\textsf{d}_i)),
\end{eqnarray*}
where the last term is obtained as a summation with respect to saddle points, labelled by $i$, and identification between $\nu$ and $m$ is assumed. The previous $f_i(k)$  which appeared in (\ref{cor3}) is obtained as $ik\textsf{d}_i$. Hence we induce a geodesic approximation (\ref{cor3}) from the assumption of the identification (\ref{ast}).

\section{Geodesic approximation and Saddle point approximation of the Green function II}

Since (\ref{ast}) is a complex relation, the correspondence we obtained in the last subsection is only a formal correspondence  unless a proper analytic continuation process is provided. We have already seen that the Green function can be analytically extended by specifying its branch cuts. We will provide an analytic continuation for the geodesic approximation and hence make the formula (\ref{cor3}) rigorous. There are two steps involved to define an analytic continuation for the functions $\textsf{t, l, d}$. The first step is to analytically continue $r_c(u)$ to an analytic function on the complex $u$-plane, and the second step is to deform the contour of integration to the complex $r$-plane. 

\begin{figure}[hpt]
\centering
\subfigure[]
{  
   \includegraphics[width=2.5in]{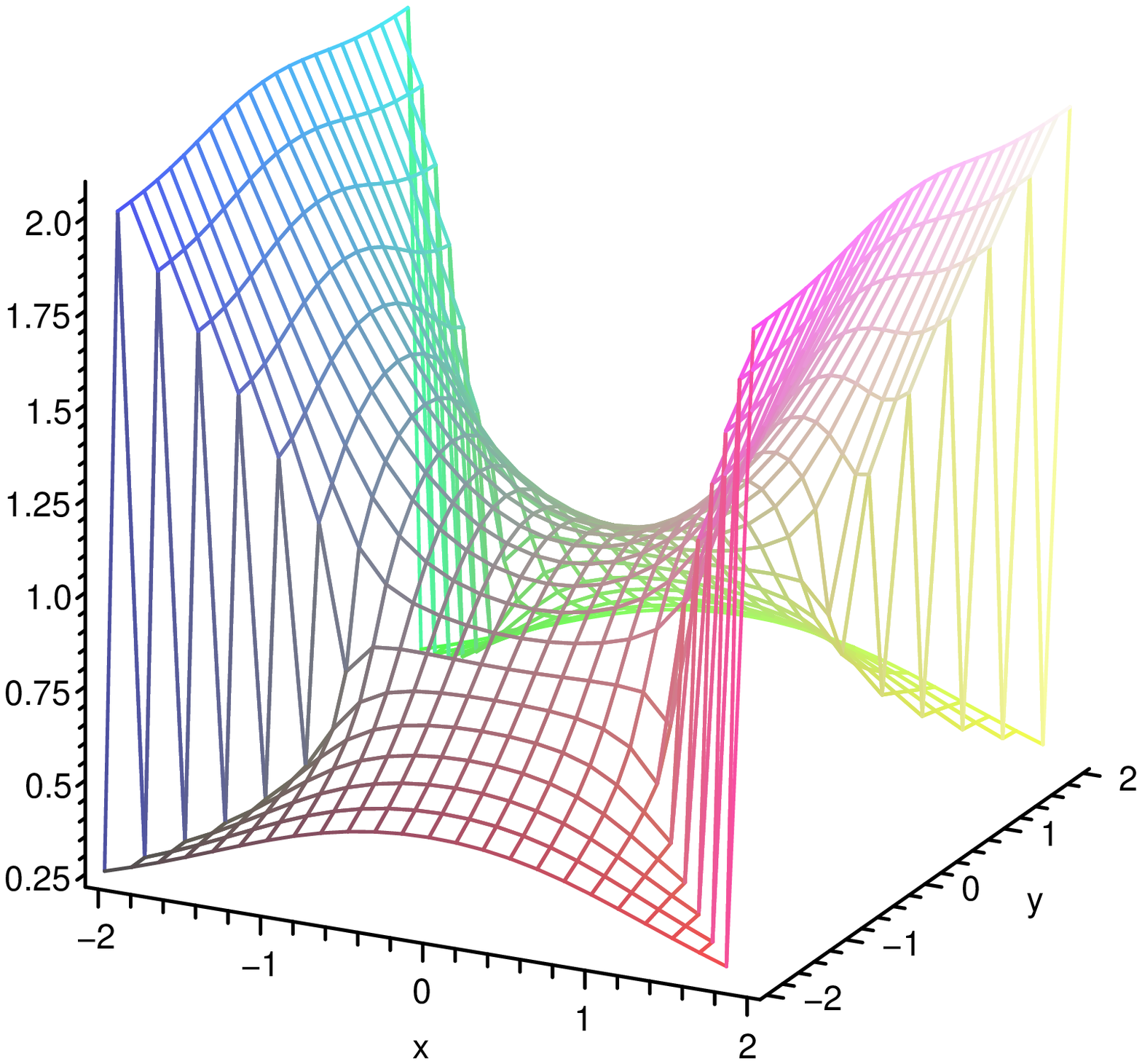}
}
\subfigure[]
{
\includegraphics[width=2.5in]{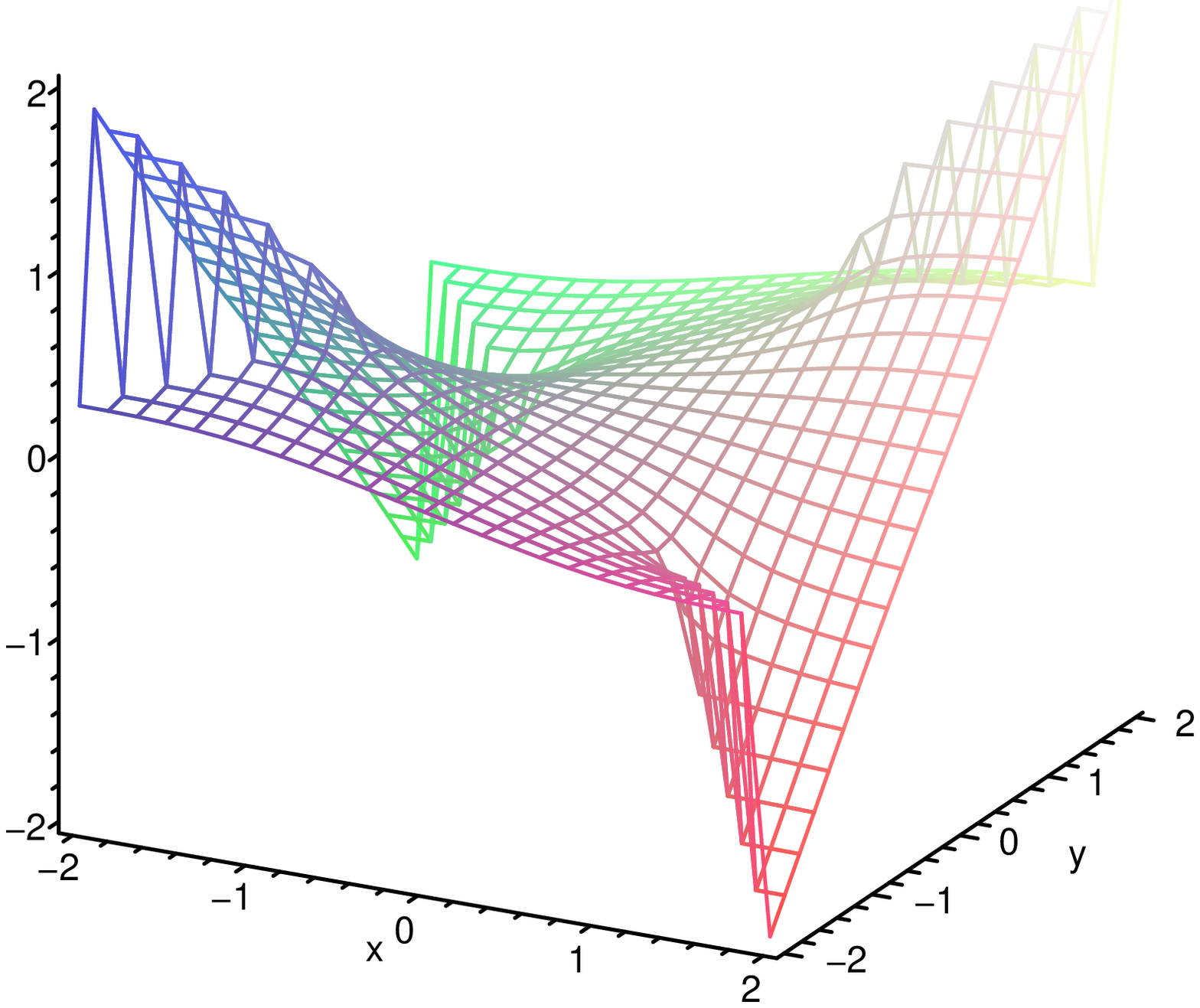}
}
\caption[]{Fig. (a) is the real part  of $r_c(u)$ on the complex $u=x+iy$-plane  when $k=1$. Fig. (b) is the imaginary part of $r_c(u)$ on the complex $u$-plane when $k=1$.}\label{fit6}
\end{figure}

For the first step, $r_c$ is a solution of (\ref{turningpt}), or equivalently,
$r(u)^4+(k^2-u^2-1)r(u)^2-k^2=0$.
If the discriminant
$
(k^2-u^2-1)^2+4k^2=0,
$
then $u$ is in the set
$$\{u=\pm k\pm i\} \cup\{u=\pm k\mp i\}.$$
Observe that these points are located at the four end points of branch cuts of momentum space Green function  at the large $\nu$ limit as in Fig. \ref{fit16}. In this case,
$r(u)$ has a root of multiplicity $4$. While the discriminant is not zero, there are four distinct roots. We will only concentrate on the one given as,
\begin{equation}
\label{ri}
r_c(u)=+\sqrt{\frac{1}{2}\left(-k^2+u^2+1+\sqrt{(k^2-u^2-1)^2+4k^2}\right)}.
\end{equation}
It can be analytically continued to the complex $u$-plane if we cut the points where the discriminant is zero. 
Since  properties we are concerned with do not differ much qualitatively  when $k$  varies as a positive real number, we will fix $k=1$ from now on. We plot the real part and the imaginary part of the analytically continued function of (\ref{ri}) in Fig. \ref{fit6}.

We want to mention some properties of the analytic function $r_c(u)$.
When $u$ approaches imaginary infinity, the norm of  the turning point $|r_c(u)|$ approaches $0$ and hence so does $r_c(u)$.  As $u$ approaches real infinity, the norm of the turning point approaches  positive infinity. As $u$ approaches  $0$ from any direction, the norm $|r_c|$ approaches one. $u=0$  corresponds exactly to our reference geodesic $G_0$. In particular, as $u$ approaches $0$ in the direction along either the real axis or the imaginary axis, the imaginary part of $r_c(u)$ vanishes. Hence the analytic function $r_c(u)$ approaches  the horizon at $r=1$ in these two cases. 

To understand the function $r_c(u)$ mathematically is quite simple. As the way we obtained, it is a holomorphic function coming from an analytic continuation of some function on the real line. Certain limiting behaviours mentioned above are quite neat. Meanwhile its physical explanation is not as clear as its mathematical explanation. The first problem which occurs is that the concept of turning point loses its usual physical meaning. For example, with a particular turning point of a complex value, we don't know how to compare it with the horizon at $r=1$. Take its norm? Take its real part or anything else? So when we see this analytic function $r_c(u)$ we keep in mind that it is a nice function coming from a real function, which has physical meaning. If we want to use it, we use it in a way related  to the original real function. Now consider the complex $u$-plane, the complex $r$-plane and the analytic function $r_c$ which maps $u$ to $r$. Physically, there are two distinct positions on the $r$-plane: the point $r=1$ is the event horizon and the origin $r=0$ is the singularity of the black hole. The previous observation concerning the function $r_c(u)$ tells us that it approaches $r=1$ when $u$ approaches zero along any direction. Therefore, we consider simply all straight lines,  $u=|u|\exp(i\theta)$ with some constant $\theta$, through the origin on the $u$-plane and find their images under $r_c(u)$ on the complex $r$-plane. 

\begin{figure}[hpt]
\centering
\subfigure[]
{  
   \includegraphics[width=2.5in]{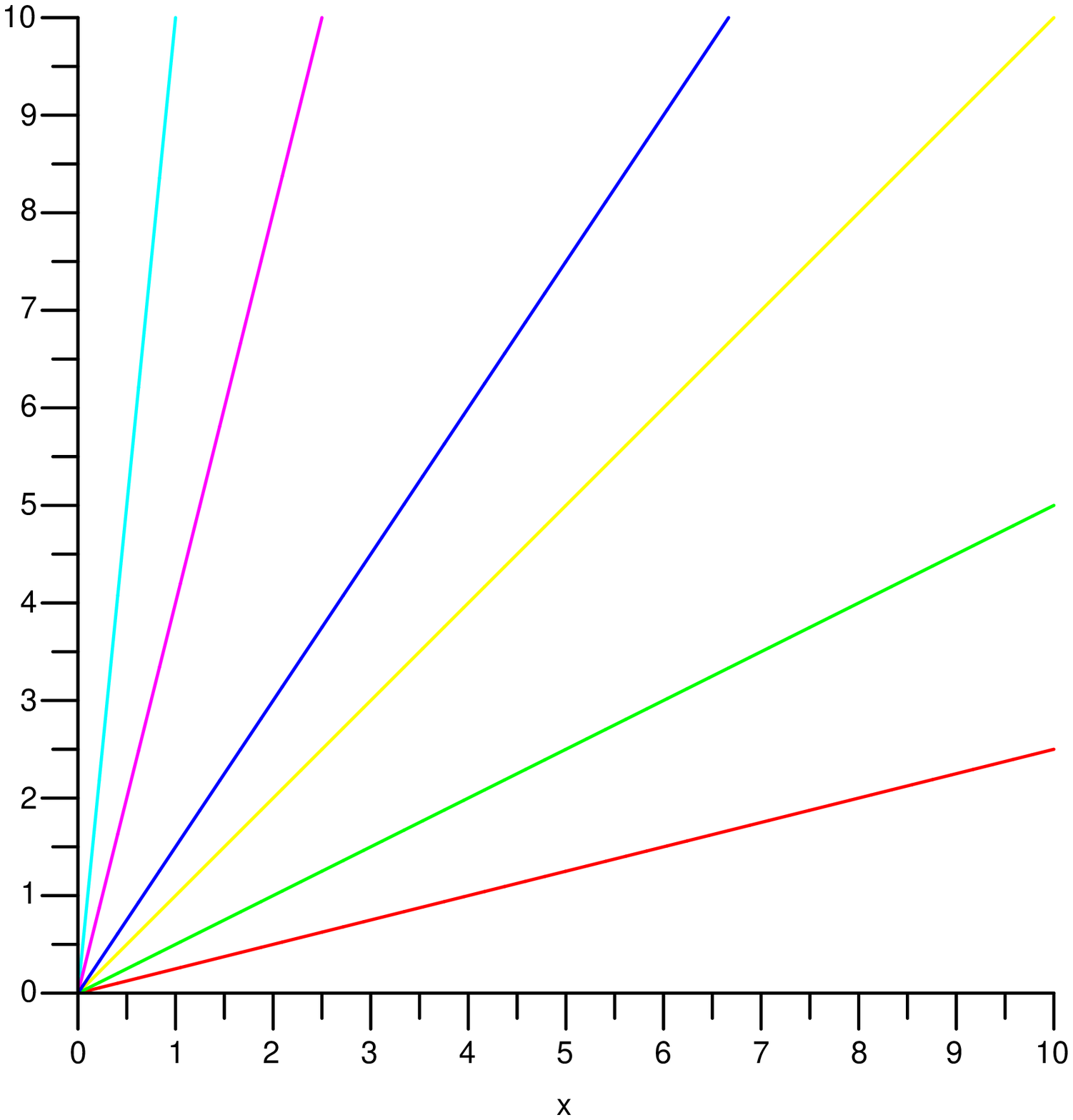}
}
\subfigure[]
{
\includegraphics[width=2.5in]{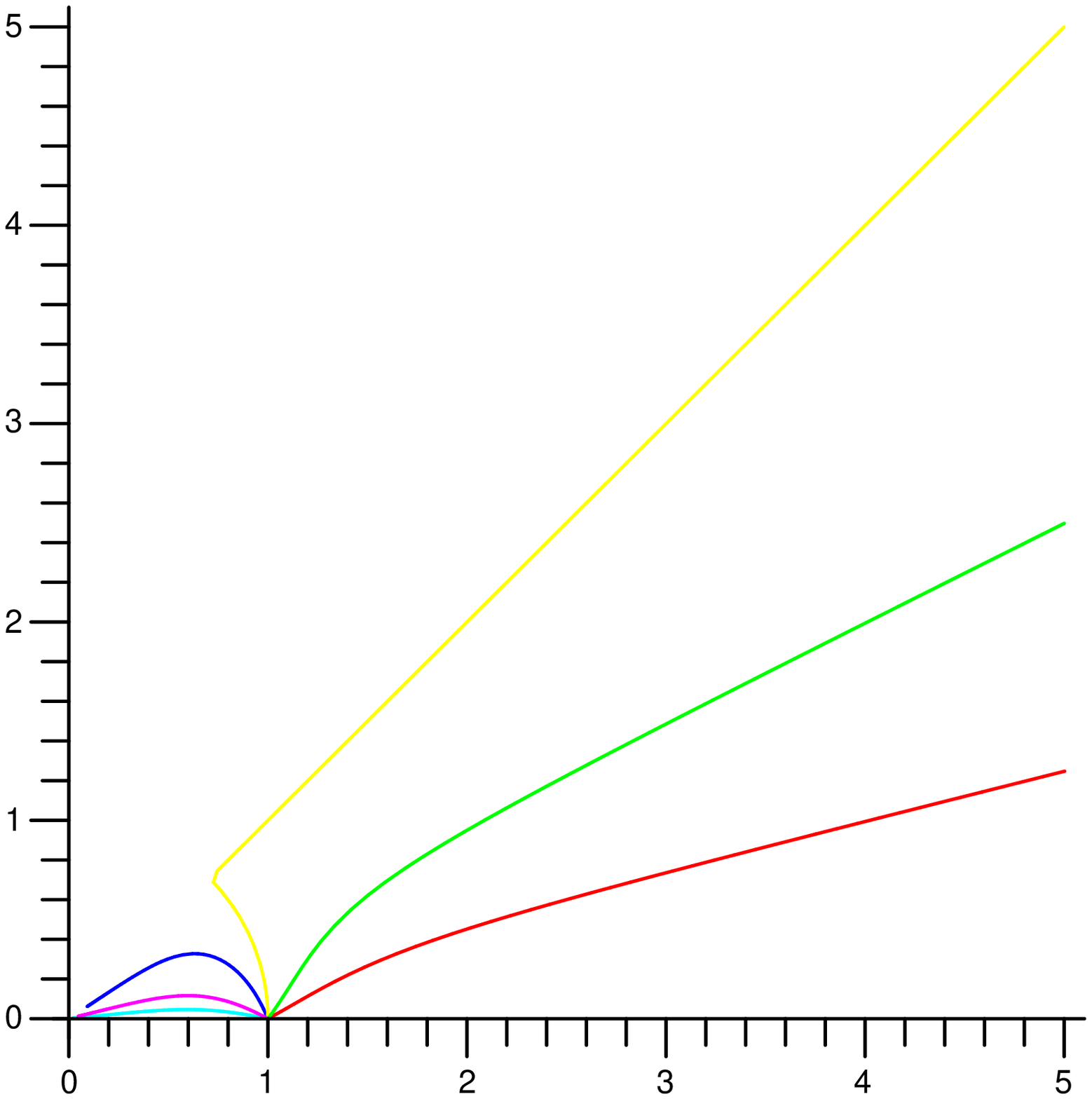}
}
\caption[]{(a) shows straight lines $u=|u|\exp(i\theta)$ with varies slopes. (b) shows  the images of $r_c(u)$ of the  lines in (a). Lines and their images are of the same colour.}
\label{uplane}\end{figure}
Particular cases of this mapping are plotted in Fig. \ref{uplane}. 

 Consider points on the line $u=|u|\exp(i\theta)$ starting from zero to the complex infinity. The analytic function $r_c(u)$ thus gives us a family of geodesics parametrized by their turning point $r_c(u)$ as $u$ moves. No matter what $\theta$ is, the starting geodesic is the one with turning point at the horizon, which is $G_0$. Therefore, we find a way to put any geodesic with an arbitrarily complex $r_c(u)$ into a family of geodesics by joining the origin to the $u$. This family is considered to consist of deformation of the reference  geodesic $G_0$. When $\theta=0$, $u$ is always real, and the family of geodesics are physically observable geodesics with real turning points starting from the horizon ($G_0$) to positive spatial infinity. When $\theta=\pi/2$, the line parametrizes a family of geodesics with all real valued turning points, starting from  horizon ($G_0$)and decrease monotonically to the spatial singularity at $r=0$. Incidentally any line with $\theta$ greater than $\pi/4$ (this particular angle actually depend on our assumption of $k=1$, other $k$ will give other angle) has turning point $r_c(u)=0$ as $u$ goes to infinity. (Fig. \ref{uplane})

\begin{figure}[hpt]
\centering
\includegraphics[width=4in]{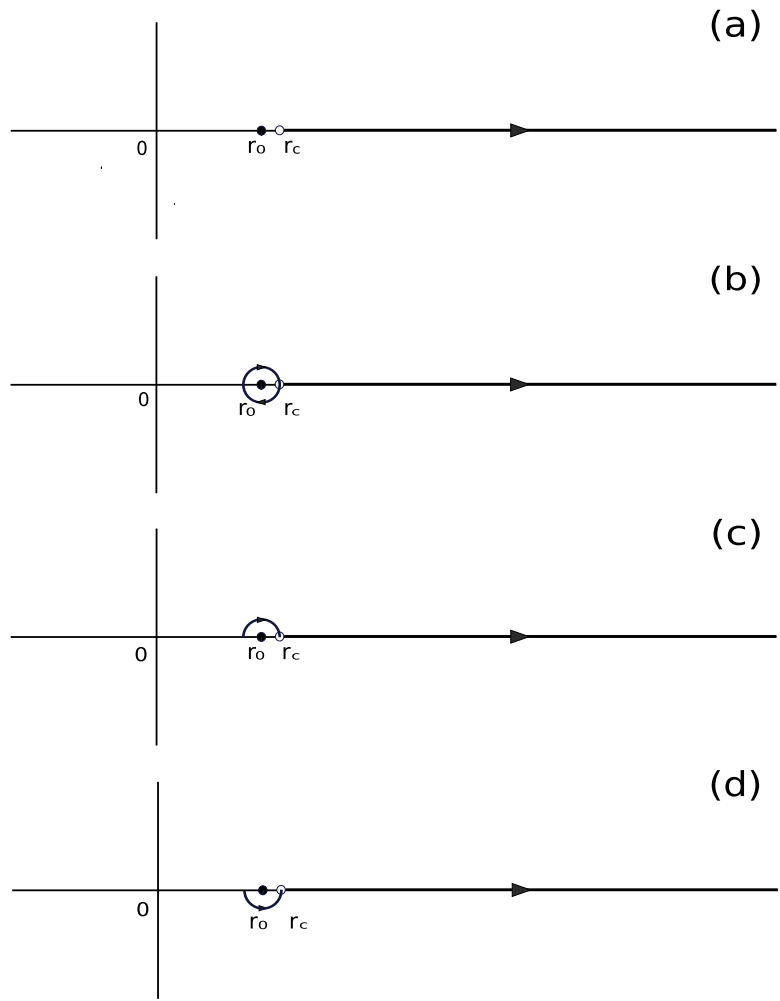}
\caption[]{Choices of deformed contours. (a) The contour chosen for $u$ approaches to $0^+$ along the real axis. (b) The contour chosen for $u$ approaches to $0^-$ along the real axis. (c) The contour chosen for $u$ approaches to $i0^+$ along the imaginary axis. (d) The contour chosen for $u$ approaches to $i0^-$ along the imaginary axis. }\label{fit17}
\end{figure}

As to the second step, the original contour in the integrations $\textsf{t}, \textsf{l}, \textsf{d}$ as in (\ref{geo3-1}), (\ref{geo3-2}), (\ref{geo3-3})  starts from $r_c(u)$ and goes to infinity along the real $r$-axis. As $r_c(u)$ is analytically continued into the complex $r$-plane,
 we need to specify the contour of integration. Although the deformation of the contour can be arbitrary, we would like to make a choice so that the heuristic relation (\ref{cor3}) gives a  proper geodesic approximation of the correlator. 
Since when $r_c(u)$ is very far from the horizon, the choice of contour is not  easy to make, we will start by considering when $r_c(u)$  gets close to the horizon at $r=1$. Write $r_c(u)=1+\epsilon \exp(i\eta)$ where $\epsilon$ is a sufficient small positive number and $\eta$ ranges from $0$ to $2\pi$. Recall  that $r_c(u)$ and $u=|u|\exp(i\theta)$ are in one-one correspondence as shown in Fig. \ref{fit6} or Fig. \ref{uplane}.
We will decide our contours according to the way how $r_c(u)$ approach to $1$ with respect to $\eta$ and hence to  $\theta$. 
As $u$ approaches to $0^+$ from $+\infty$ along the real axis ($\theta=0$), they are proposed to relate to geodesics with difference in proper time from the geodesic $G_0$ only  in the Euclidean section. Their turning points  vary from $r=+\infty$ to $r=1$. For example, geodesics $A$, $B$ and $C$ plotted in (b) of Fig. \ref{fit11}. This is actually the case before analytical continuation, so we simply choose the original contour as in (a) of
 Fig. \ref{fit17}. 
In this case, we can see that the pole at horizon does not contribute. 
 This immediately implies the choice of the contour for $u$ approaches to $0^-$ from $-\infty$ along the real axis ($\theta=\pi$) as in Fig. \ref{fit17} (b). Indeed, simply consider a closed contour in (a) of Fig. \ref{fit11}: it starts at $\tau=0$, $r=\infty$, and then goes  very close to the horizon and finally goes to $r=\infty$ ( as $C$). After this, it moves downwards along  the arc  a little bit and comes back along the dashed geodesic, which is very close to the horizon. Finally it goes back to the point where it starts. Integration along this closed contour will give proper time difference by $2\pi i$. While we have already chosen the contour as (a) of Fig. \ref{fit17} so that the outwards geodesic avoids the singularity, we have to make a choice so that the backwards geodesic can pick up the factor of $2\pi i$. The contour in  (b) of Fig. \ref{fit17}, when circling around the pole,  picks up this factor. Hence the choice is confirmed. 

That leaves us two more directions to consider.
For the purpose of making a geodesic approximation, we are interested in geodesics joining two boundaries as $A$, $B$, $C(G_0)$ and $D$ in the Penrose diagram in Fig. \ref{fit11}. These geodesics correspond to $u=iu_I$, since there are no Euclidean time difference between them and the reference geodesic $G_0$. The imaginary part  $u_I$ is indeed related to the Lorentzian proper time of the geodesics, since the Lorentzian time separation of $G_0$ is zero. We want to make a choice of contour so that the jumping in the Euclidean time $-i\beta/2$ is included even if we only carry out the integration for the Lorentzian section. As $u_I$ moves from  $+i\infty$ to $0$ along the imaginary axis, $r_c(u_I)$ moves from $r=0$ to the horizon  along the real axis. As $r_c(u_I)$ goes very closed to $r=1$, we choose the contour to be (c) of Fig. \ref{fit17} so that this naturally include the jumping of $-i\beta/2$ by making use of the pole. Symmetrically, as $u_I$ moves from $-i\infty$ to $i0^-$ along the imaginary axis, the contour is chosen as (d) of Fig. \ref{fit17}.

Therefore, for the turning point very closed to the horizon, we have made our choices for $\theta=0, \pi/2, \pi, 3\pi/2$ as in (a), (c), (b), (d) in Fig. \ref{fit17}, respectively. We make the following requirement for the rest of the $\theta$'s: when $0\leq \theta<\pi/2$, the contour is chosen to be as in (a); when $\pi/2\leq\theta<\pi$, the contour is chosen to be as in (c); when $\pi\leq\theta<3\pi/2$, the contour is chosen to be as in (b); when $3\pi/2\leq\theta<2\pi$, the contour is chosen to be as in (d). As to $u$ away from $0$, i.e. $|u|$ not very small,  $r_c(u)$ is far from the horizon. We simply choose the contour according to the argument $\theta$ of $u$. 

Combining the two steps, we obtain a complete  analytic continuation procedure. In this way the previous  heuristic relation (\ref{cor3}) is exactly defined. A interesting phenomenon is that the Lorentzian section of geodesics, which is related to the imaginary part of the saddle points, can have turning point inside horizon. In this sense, we may be able to  find a signature of the BTZ singularity from the boundary correlation function.
However, we by no means imply this is the only possible geodesic approximation. It is possible to assign different correspondence for $u$ and $r_c$ firstly, then to define a corresponding analytic continuation according to its relation with the correlation functions secondly. That is to say, it is possible to define a different way of geodesic approximation for the same correlation function. 
Similar ideas can be seen in \cite{kraus-2003-67}, in which two different but equivalent analytic continuations for position space Green function are presented.   

\section{Application of the frequency-geodesic relation}
With the exact geodesic approximation provided in the last section, we would like to see how it works  through a simple computation and its physical implication. 
We will first look for saddle points of the correlation function (\ref{twopt}) and then look for their corresponding geodesics. There are two cases of interest. The first case is when the time separation is purely Euclidean as in (b) of Fig. \ref{fit11}. The second case is when the two points sit on different boundaries in the Lorentzian section of space time  with time separation $\Delta t-i\beta/2$ where $\Delta t$ is the Lorentzian time separation as in (a) of Fig. \ref{fit11}. 

In the large $\nu$ limit, instead of finding $S_0(z)$ and hence $Z_0(u, k)$ as we did before, we adopt a slightly different procedure. Since we already have the explicit expression of the momentum space correlation function, we obtain the approximation simply by considering its asymptotic form in the large $\nu$ limit. 
After rescaling $\omega, J$ to $u, k$, we can  write (\ref{g12}) in the form of (\ref{twopt}) and get the expression of $2Z(u, k)=1/\nu\ln \tilde{\mathcal G}_{\beta}{}^{+}(u, k)$ as
\begin{eqnarray*}
2Z(u, k)
&=&3\pi u+\frac{1}{\nu}\ln\left(\frac{\nu u}{2\pi\Gamma(1+\nu)^2}\right)
+\frac{1}{\nu}\ln\left(\Gamma\left(\frac{\nu}{2}\left(iu+ik+1+\frac{1}{\nu}\right)\right)\right)\\
&&+\frac{1}{\nu}\ln\left(\Gamma\left(\frac{\nu}{2}\left(iu-ik+1+\frac{1}{\nu}\right)\right)\right)\\
&&+\frac{1}{\nu}\ln\left(\Gamma\left(\frac{\nu}{2}\left(-iu+ik+1+\frac{1}{\nu}\right)\right)\right)\\
&&+\frac{1}{\nu}\ln\left(\Gamma\left(\frac{\nu}{2}\left(-iu-ik+1+\frac{1}{\nu}\right)\right)\right).
\end{eqnarray*}
Applying the  Stirling formula \cite{Abramowitz} to the Gamma functions appearing %
 in $2Z(u, k)$, we get
\begin{eqnarray*}
2Z(u, k)&=&\frac{1}{\nu}\left(\ln\left(\frac{\nu}{2\pi\Gamma(1+\nu)^2}\right)+2\ln 2\pi-2\nu-2+2\nu\ln\left(\frac{\nu}{2}\right)\right)+\frac{1}{\nu}\ln(u)\\
&&+3\pi u+\frac{1}{2}(iu+ik+1)\ln(iu+ik+1)\\&&+\frac{1}{2}(iu-ik+1)\ln(iu-ik+1)+\frac{1}{2}(-iu+ik+1)\ln(-iu+ik+1)\\
&&+\frac{1}{2}(-iu-ik+1)\ln(-iu-ik+1).
\end{eqnarray*}
The derivative with respect to $u$ and is then obtained as follows,
\begin{eqnarray}
\label{dzu}
\frac{2dZ(u, k)}{du}&=&
3\pi+\frac{i}{2}\ln{\frac{(iu+ik+1)(iu-ik+1)}{(-iu+ik+1)(-iu-ik+1)}}, \qquad\text{ as }\nu>>1.
\end{eqnarray}
 To carry out the steepest descent method, we may write (\ref{twopt}) in  the form of (\ref{form}). In this way, $g(u)$ will be simply the constant $\nu$ and 
$$h(u)=-iu\left(\Delta t-i\frac{\beta}{2}\right)+2Z(u, k).$$
Saddles of $\tilde G_{12}(\Delta t -i\beta/2, k)$ are thus determined by $dh(u)/du=0$, that is,
\begin{equation}
\label{saddles}
\frac{2dZ}{du}=i\Delta t+\frac{\beta}{2}
\end{equation}
We write $\Delta t=t+i\tau$ for $t$ and $\tau$ both real numbers, i.e. $t$ is the time separation in Lorentzian section and $\tau$ is the time separation in the Euclidean section. We can see that $t=0$ is our first case where the time separation is purely imaginary. 
 $\tau=0$ corresponds to the second case.
 From (\ref{dzu}) and (\ref{saddles}), we can find saddle points by solving the following equation,
$$\frac{(iu+ik+1)(iu-ik+1)}{(-iu+ik+1)(-iu-ik+1)}=\exp(2(t+i\tau)),$$
where  the inverse Hawking temperature $\beta=2\pi$ is substituted in.
There are two solutions,
$$u_{\pm}(\Delta t)=\frac{i(1+ \exp(2\Delta t))\pm\sqrt{-4 \exp(2\Delta t) +k^2(1-\exp(2\Delta t))^2}}{1-\exp(2\Delta t)}.$$
For simplicity, we will again assume $k=1$. Computation can be carried out for other non-zero positive $k$ similarly. 
When $t=0$, we can view saddles as being parametrized by the Euclidean time  $\tau$. Since the Euclidean time separation $\tau$ is of period $2\pi$, we may assume $\tau\in(-\pi, \pi]$. The saddles are
\begin{equation}
\label{eusaddle}
u^E_{\pm}(\tau)=\frac{i(1+ \exp(2i\tau))\pm\sqrt{-4 \exp(2i\tau) +(1-\exp(2i\tau))^2}}{1-\exp(2i\tau)}.
\end{equation}

We observe that $u^E_{\pm}(\tau)$  have vanishing imaginary part. This implies that the time separation is purely Euclidean, the possible dominating geodesics of the correlation function will not have Lorentzian time section. Another observation from the solutions is that the two saddles are both periodic functions with respect to $\tau$  of period $\pi$.
In particular, $u^{E}_-(\tau-\pi)=u^E_+(\tau)$. So there is in fact only one solution 
\begin{figure}[hpt]
\centering
\includegraphics[width=3in]{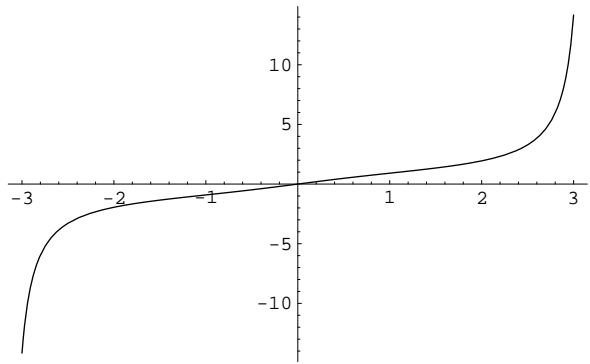}
\caption[]{$u^E_{\pm}$ parametrised by Euclidean time $\tau$. As $\tau$ varies from $-\pi$ to $0$, the graph of $u^E_-(\tau)$ is shown. As $\tau$ varies from $0$ to $\pi$, the graph of $u^E_+(\tau)$ is shown.}
\label{fit2}
\end{figure}
as seen in Fig. \ref{fit2}.

\begin{figure}[hpt]
\centering
\subfigure[]
{
   \includegraphics[width=2in]{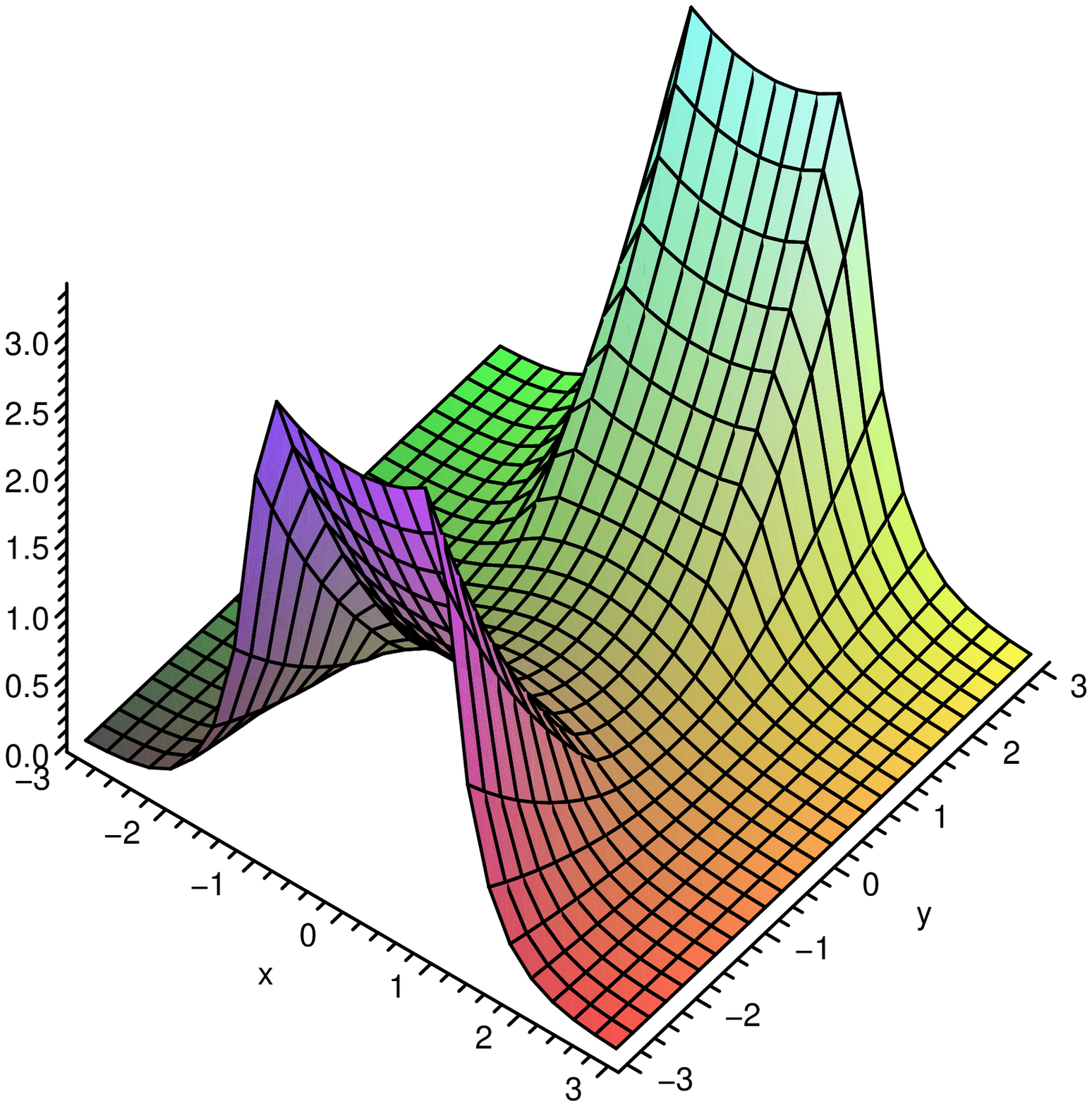}
}
\hspace{0.6cm}
\subfigure[]
{  
   \includegraphics[width=2in]{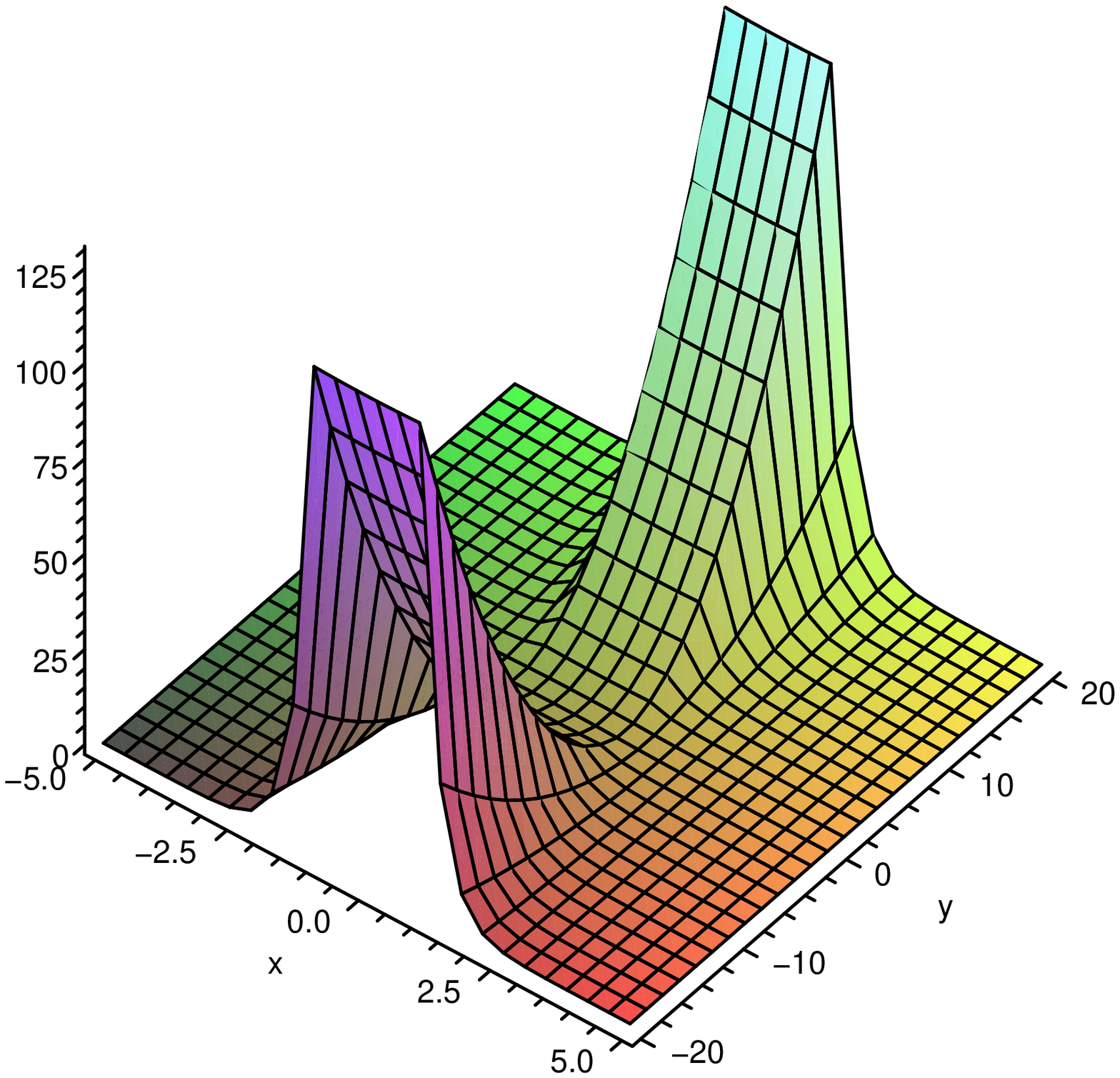}
}
\subfigure[]
{
\includegraphics[width=2in]{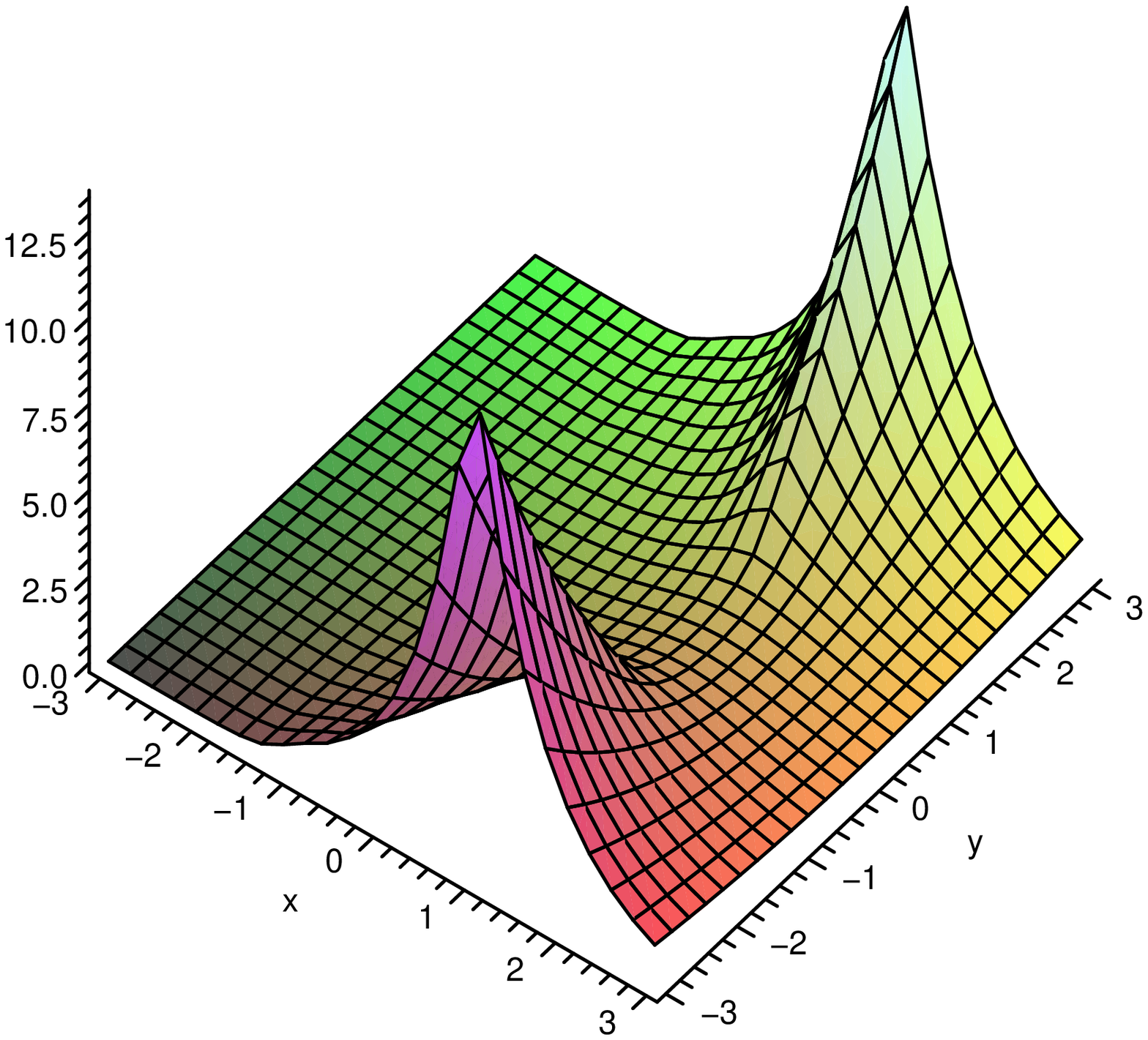}
}
\hspace{0.6cm}
\subfigure[]
{
\includegraphics[width=2in]{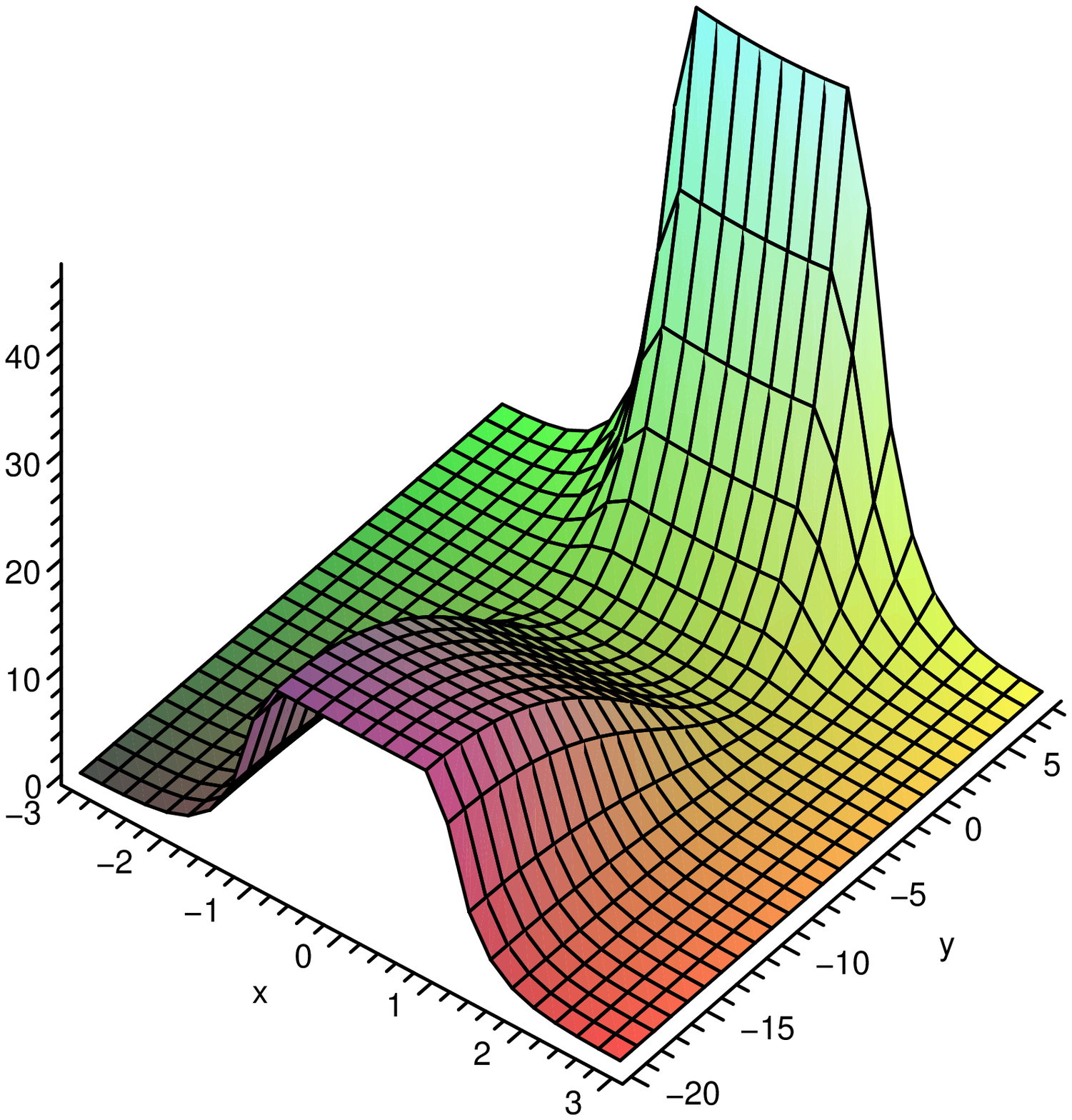}
}
\subfigure[]
{
\includegraphics[width=2in]{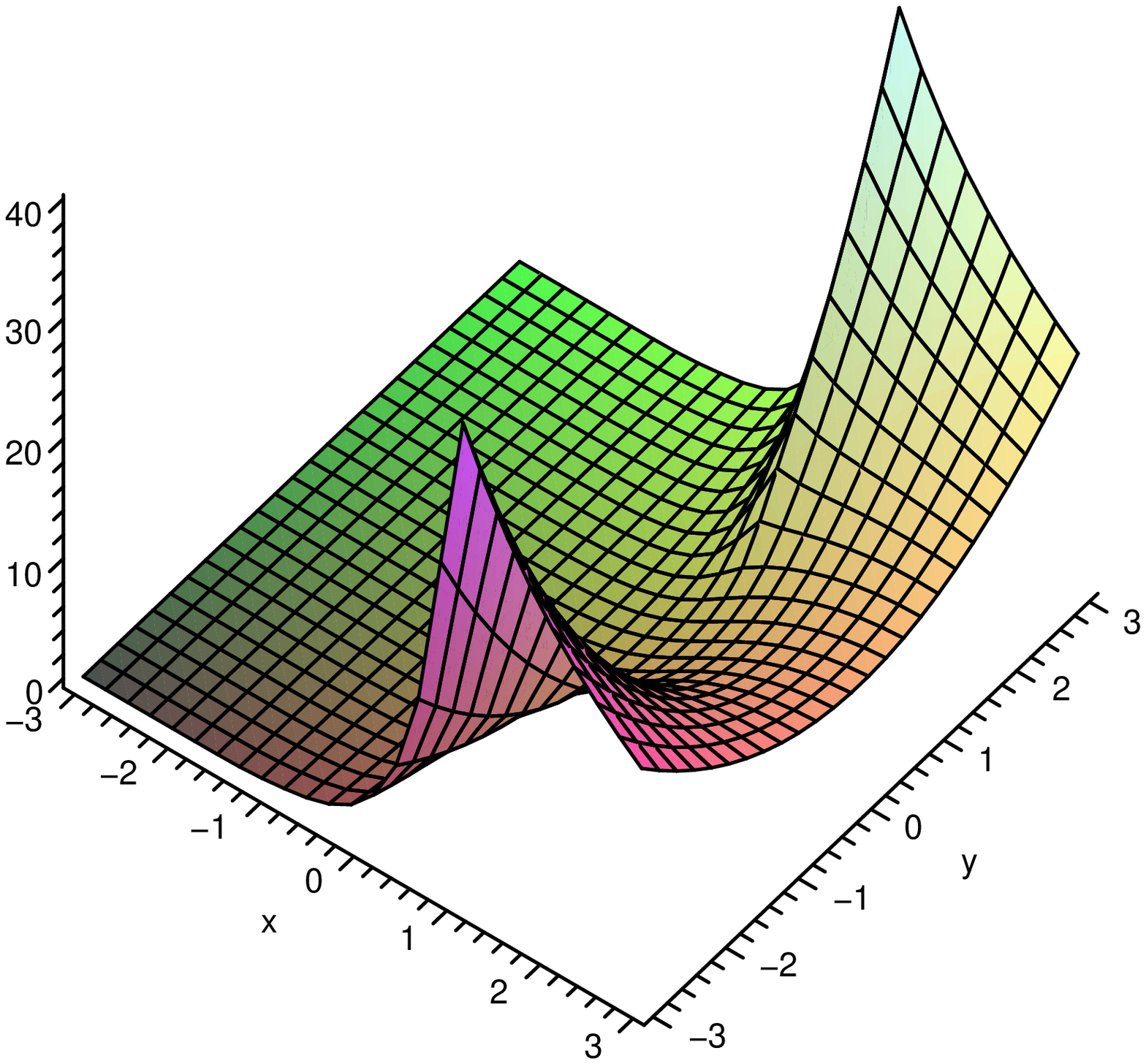}
}
\hspace{0.6cm}
\subfigure[]
{
\includegraphics[width=2in]{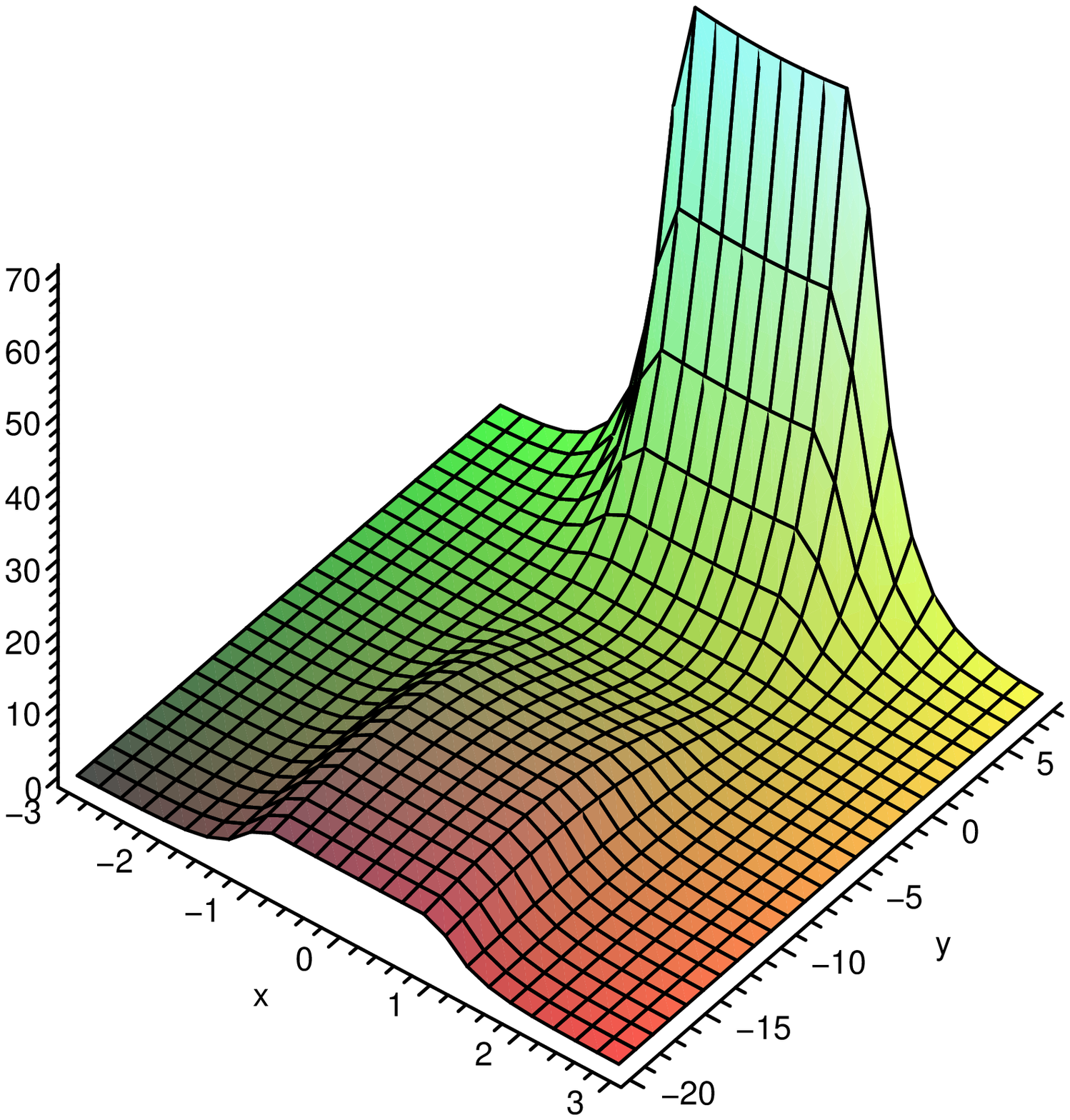}
}
\caption[]{Graphs of $|\exp(h(u))|$ on the complex $u$-plane. (a), (c) and (e) are  parametrized by the Euclidean time separation $\tau$ when $t=0$. (a) is when $\tau=0$, the saddle is at $u=(0, 0)$. (c) is when $\tau=1.5$. (e) is when $\tau=2.5$. (b), (d) and (f) are  parametrized by the Lorentzian time separation $t$ when $\tau=0$. (b) is when $t=0$. 
(d) is when $t=0.1$, the upper saddle moves downwards and the lower one moves upwards rapidly. (f) is when $t=1.5$, the two saddles stops at the lines $x=\pm 1$. 
the lower part of the graph reduce height rapidly while the upper half increases its height rapidly.
}
\label{expg}\end{figure}

Besides the formula (\ref{eusaddle}), the behaviour of the saddle can be seen clearly from the graph of the correlation itself as in the left column of Fig. \ref{expg}. As the Euclidean time difference $\tau$ increases from $0$ to $\pi$, the saddles move from $0$ to the right along the real $u$-axis. For the limiting case when $\tau=\pi$, the saddles actually move to positive infinity. The changing of the shape of the graph is with respect to the moving of the saddle.  Symmetrically, when $\tau$ varies from $0$ to $-\pi$, the saddle starts from the origin and move to left till negative infinity along the real axis (not shown in the figure). 

By evaluating the second derivative $d^2 h(u)/du^2$ at the saddles, we find that the saddle in this case is of degree one for all $\tau\in(-\pi, \pi]$ and  the steepest descent contour is the real axis itself. Thus in this case, there is no need to deform the integration contour,  and the saddle obtained dominates the integration.



\begin{figure}[hpt]
\centering
\includegraphics[width=2.5in]{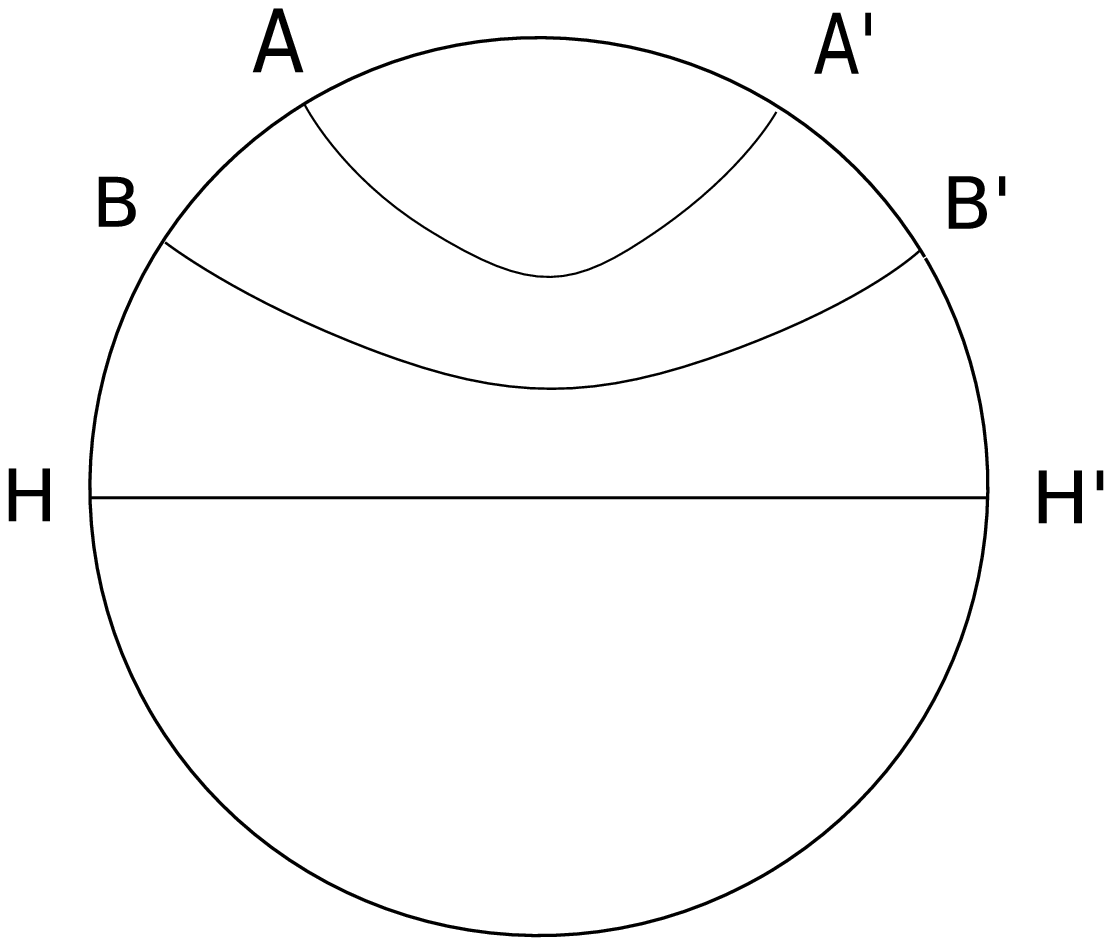}
\caption[]{As saddle moves from the origin to the positive real axis, the corresponding geodesic moves from $H'H (\text{i.e. }G_0)$ to $B'B$ and then $A'A$.}
\label{eucgeo}\end{figure}
According to the frequency-geodesic relation specified in the last section, we can find the corresponding geodesics with the saddle points. These geodesics will play the role of dominating geodesics in the geodesic approximation of the two-point function. 
 As illustrated in Fig. \ref{eucgeo}, as $u$ increases from $0$, the corresponding geodesic moves from $H'H$, to $B'B$ and then $A'A$.   As the saddle increase, the turning point of the corresponding geodesic move from the horizon to infinity. This is also what we expect.
 
\begin{figure}[htp]
\centering
{  
   \includegraphics[width=3in]{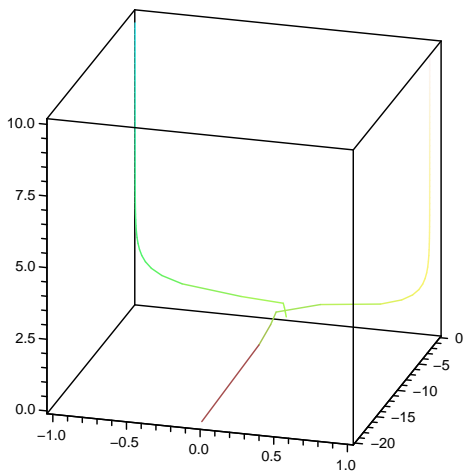}
}
{
}
\caption[]{ Two trajectories of the upper and lower saddles $u^L_{\pm}(t)$ parametrized by Lorentzian time $t$ (increases in the vertical direction) on the complex $u$-plane. 
}
\label{fit3}
\end{figure}

 The second case we want to consider is $\tau=0$. Saddles are parametrized by the Lorentzian time $t$ and given as
\begin{equation}
\label{lorsaddle}
u^L_{\pm}(t)=\frac{i(1+ \exp(2t))\pm\sqrt{-4 \exp(2t) +(1-\exp(2t))^2}}{1-\exp(2t)}.
\end{equation}
Saddles are plotted as in the Figure \ref{fit3}. 
For each time $t$ between $0$ and $\infty$ there are two saddle points. As $t$ varies from $0$ to around $0.88$, the upper saddle moves from the origin downwards to $-i$ and the lower saddle moves from negative imaginary infinity upwards to $-2i$ rapidly. The two saddles ``meet'' when the time separation is about $1.10$. After that, the upper saddle moves to the left and the lower one moves to the right, both with constant imaginary part. After the time separation exceeds $t=1.7$, the two saddles disappear into the branch cuts around $u=\pm 1-i$. 

\begin{figure}[hpt]
\centering
\subfigure[]
{  
   \includegraphics[width=1.9in]{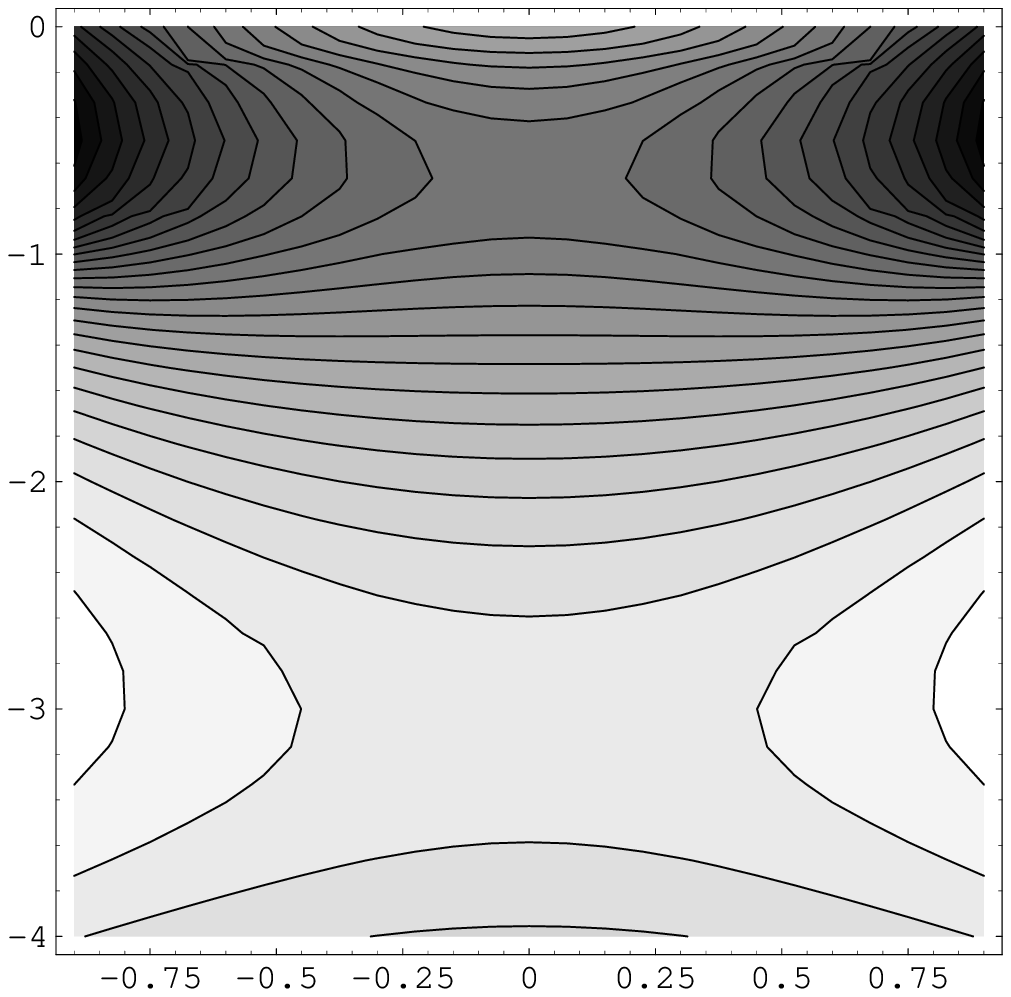}
}
\hspace{0.7cm}
\subfigure[]
{
\includegraphics[width=2.0in]{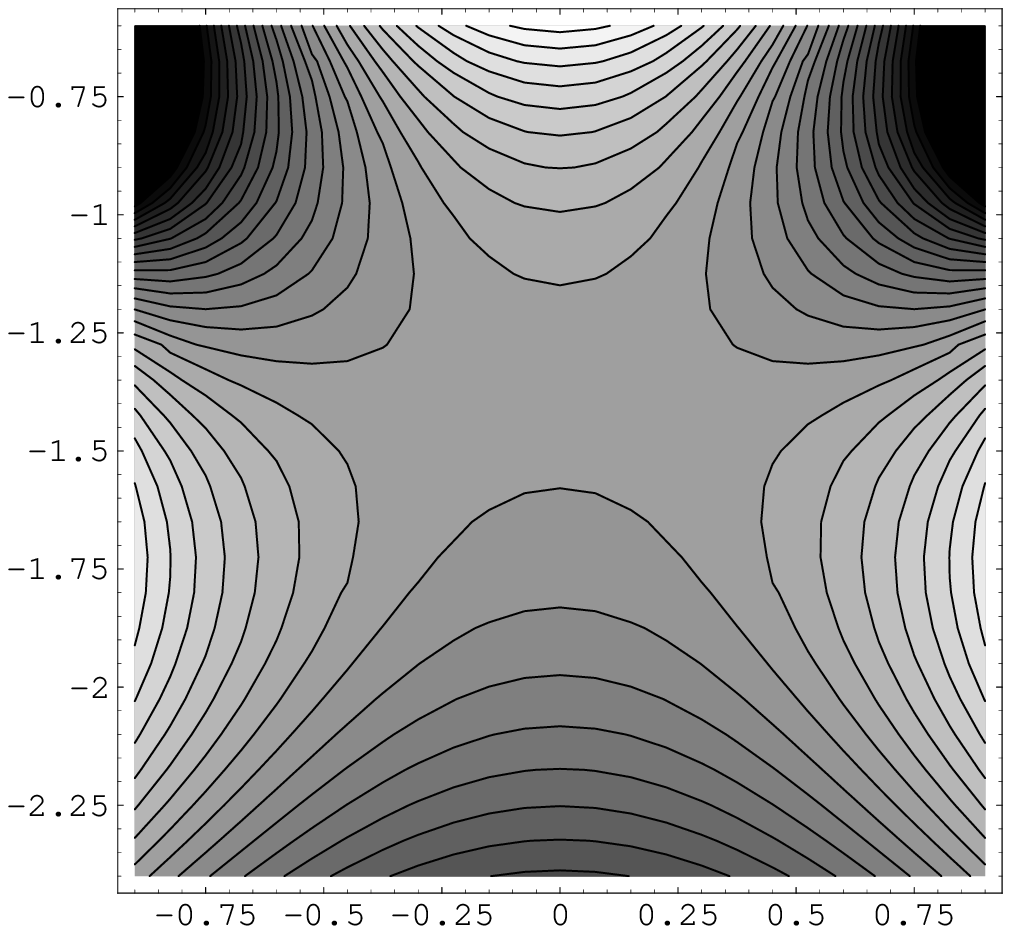}
}

\subfigure[]
{
\includegraphics[width=2.0in]{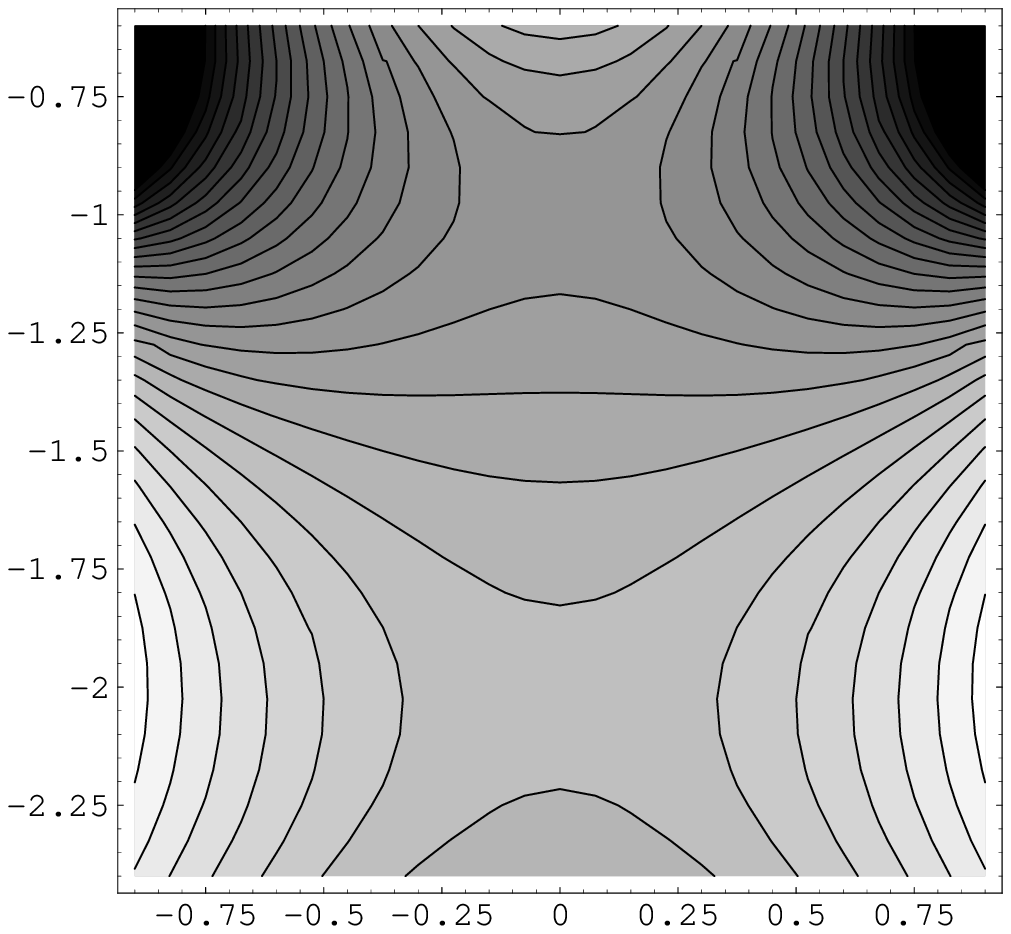}
}
\hspace{0.7cm}
\subfigure[]
{  
   \includegraphics[width=2.0in]{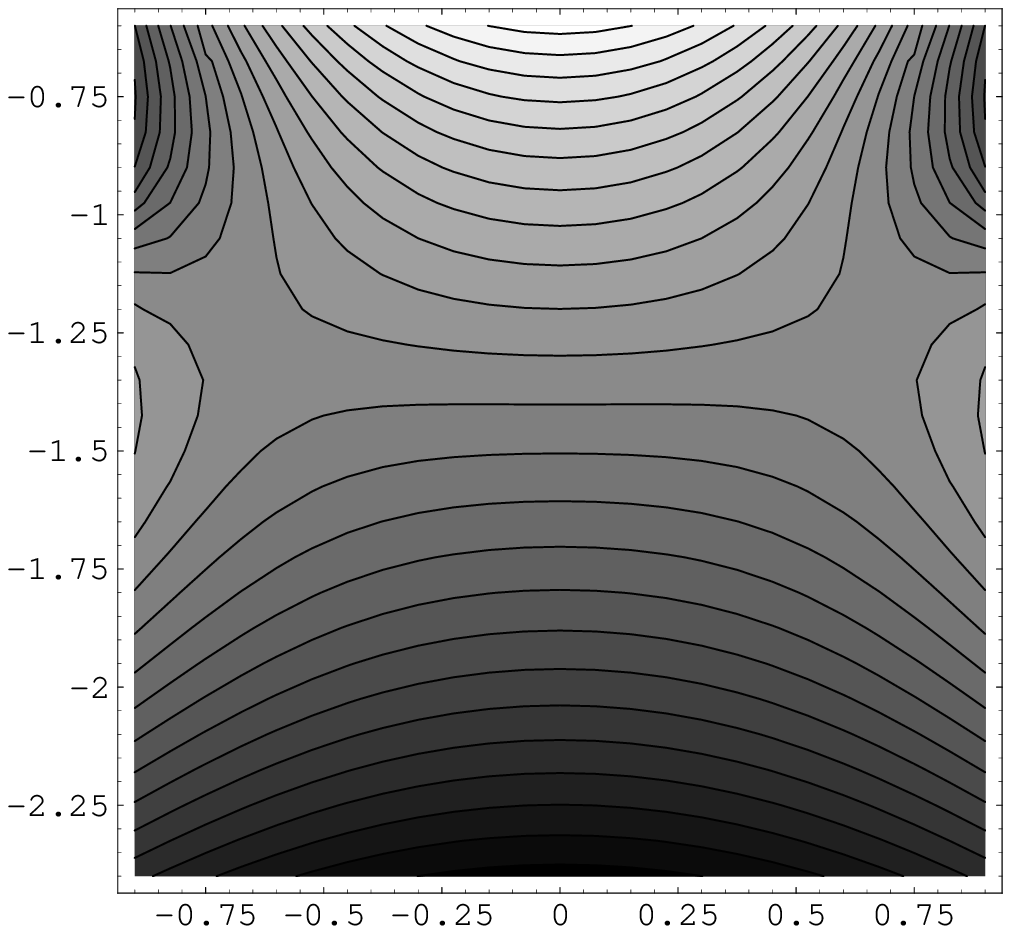}
}

\subfigure[]
{
\includegraphics[width=2.0in]{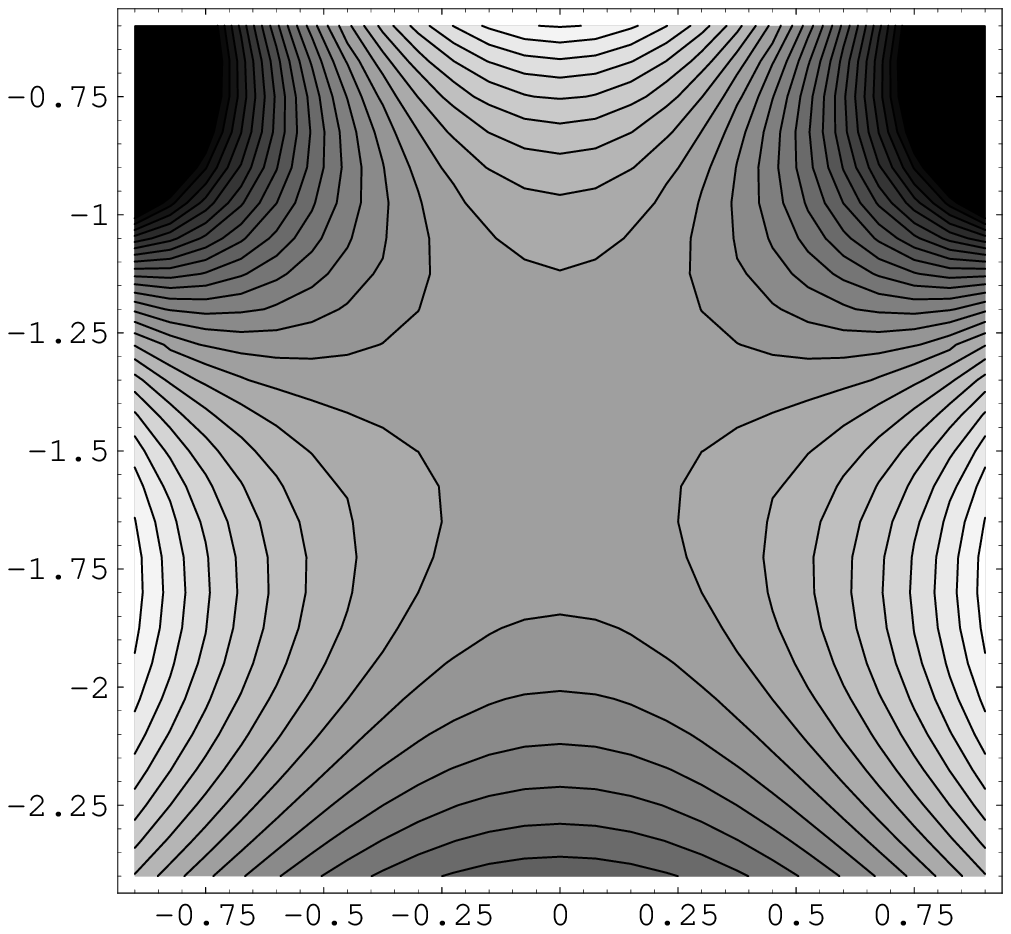}
}
\hspace{0.7cm}
\subfigure[]
{
\includegraphics[width=2.0in]{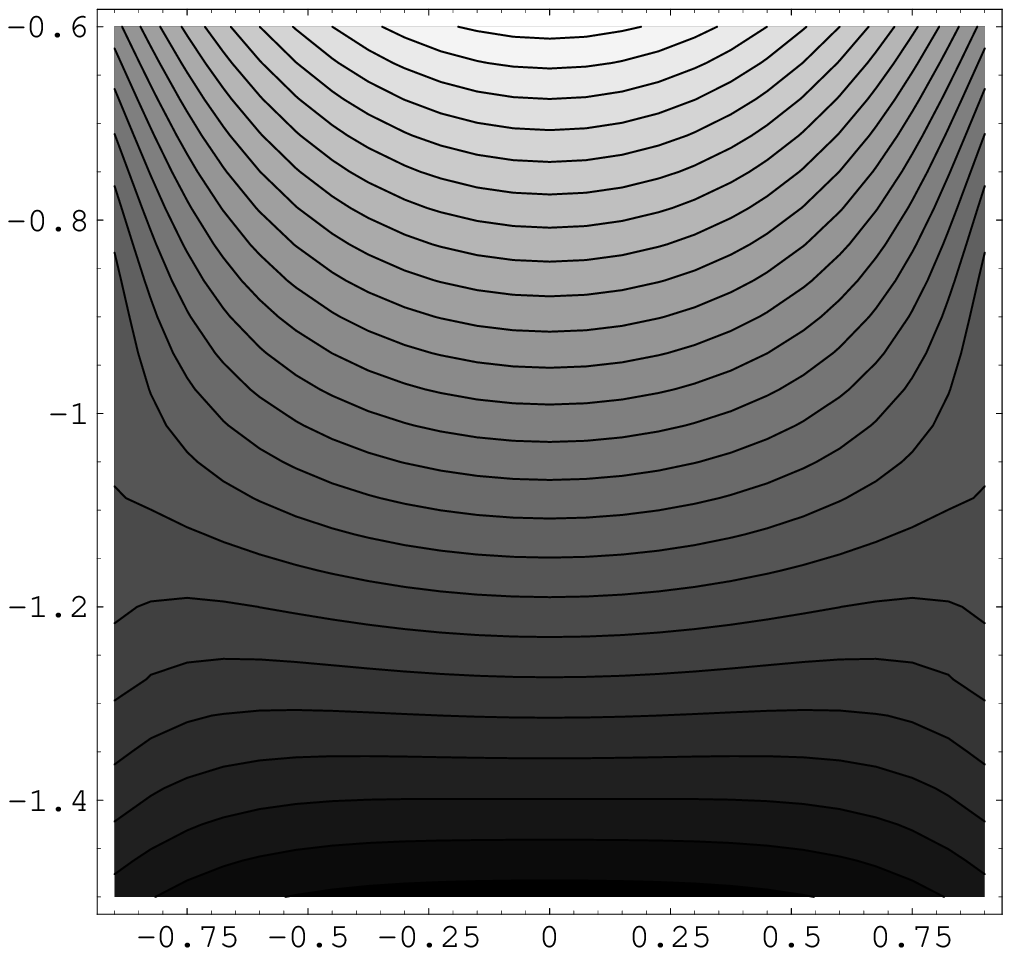}
}
\caption[]{Saddles move with respect to Lorentzian time separation $t$. (Suggest to read the left column first and then the right one.) Shading in the figure has the property:  darker means lower and lighter means higher.   (a) $t=0.60$. (c) $t=0.80$. (e) $t=0.88$. (b) $t=0.90$. (d) $t=1.10$. (f) $t=1.70$. The upper saddle dominates in (a), (c) and (e). Both saddles dominate in (d). Saddles disappear on to the branch cut in (f).}
\label{euexpg}\end{figure}
A magnified picture indicating the ``collision'' of the upper and lower saddles is shown in Fig.  \ref{euexpg}. 
The behaviour of saddles can also be seen from the graph of the correlation function itself as shown in the right column of Fig. \ref{expg}. When the Lorentzian time separation is zero, 
the upper saddle is at the origin and the lower one is at $-i\infty$.  There are two ``lumps'' (both diverge at infinity) in the graph. We refer them as the upper and lower part. As the time separation increase, the lower part descend (this corresponds to the lower saddle, who always sits on top the ``lump'', moving upwards) and the upper parts at the same time grow. When the time separation becomes sufficiently large, the lower part almost becomes flat and the upper part grows very tall. 

These are both of the saddles solved as solutions of $dh(u)/du=0$. However, not all saddle points will contribute in the final approximation. We need to find out their steepest descent direction and decide which one can be used in the approximation.
If we evaluate the second derivative $d^2 h(u)/du^2$ at the saddles obtained, we can determine the steepest descent direction for each saddle. This is also indicated in Fig. \ref{euexpg}.  We can see that before the two saddles collide, the upper saddle has steepest descent direction in the horizontal direction.  The lower saddle has its steepest descent in the vertical direction. Since the contour of the Fourier integration is along the real $u$-axis, we conclude that the dominant saddle is the upper one. In the cases when the saddles are colliding and after collision, their positions and steepest contours are symmetric with respect to the $y$ axis, so both of the saddles contribute. 

\begin{figure}[hpt]
\centering
\subfigure[]
{
   \includegraphics[width=2in]{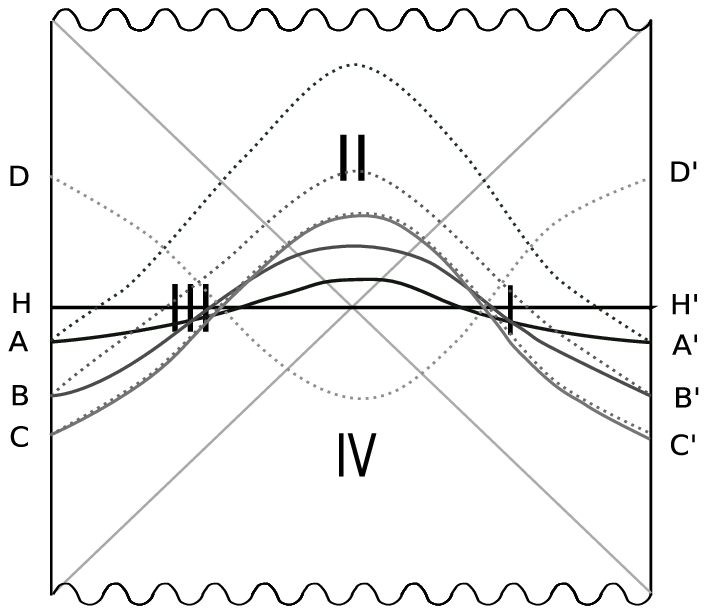}
}
\hspace{1cm}
\subfigure[]
{
\includegraphics[width=2in]{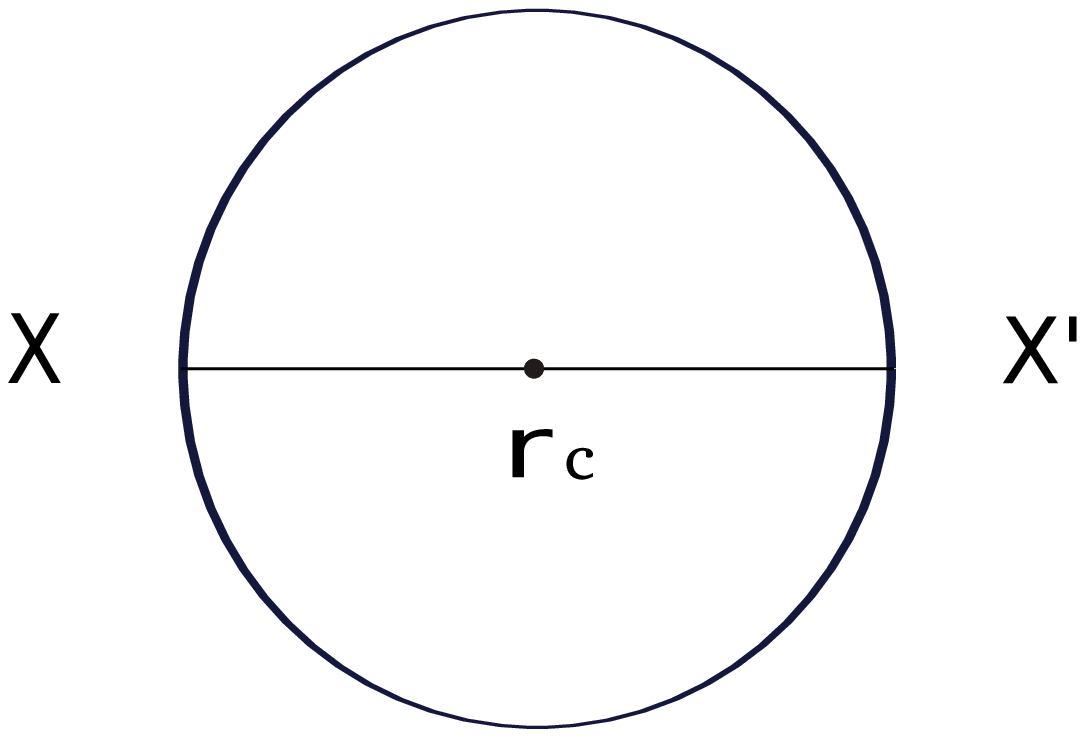}
}
\caption[]{Geodesics in the Lorentzian and Euclidean sections before  the saddle points collide. (a) The Lorentzian section. Geodesics with solid lines  correspond to the upper dominating saddle at various time separation. Geodesics with dashed lines correspond to the lower not-dominating saddle at various time separation. (b) The Euclidean section at time separation $t$. When $t=0$, $X'X=H'H$ and $r_c=1$. As $t$ increases, $X'X$ stands for $A'A$,  $B'B$ and finally $C'C$,  with $r_c$ their corresponding turning points.}
\label{lorengeo}
\end{figure}
Now we would like to see how  movements of  saddle points imply the geodesic approximation. At $t=0$, the upper saddle is at the origin $u=0$, and this corresponds to the reference geodesic $G_0$. This geodesic is shown as $HH'$  in (a) of Fig. \ref{lorengeo}. The lower saddle is at imaginary infinity and does not count as a saddle yet. When the Lorentzian time separation $t$ increases (from $H'H$, to $A'A$,  $B'B$ and finally to $C'C$ in (a) of Fig. \ref{lorengeo}), the upper saddle moves downwards; the lower saddle moves upwards. Correspondingly, for solid geodesics (related to the upper saddle, which is dominant),  their turning points move from the horizon to inside the horizon. For  dashed geodesics (related to the lower saddle, which is not dominant), their turning points come out from the singularity but remain inside the  horizon. When the Lorentzian time separation increases to $C'C$, which is related to the ``collision'' of the two saddles, the solid geodesic and the dashed one almost coincide. Consider projections of these geodesics onto the Euclidean section. Since the relating saddles both have vanishing real part, both projections will have time separation exactly $-i\beta/2$ as the reference geodesic $G_0$. Therefore, for any geodesic appearing in (a), we can plot the centre of of the circle to be its turning point $r_c$. Then the Euclidean projection of the geodesic will be  $X'X$. $X'X$ stands for  $H'H$, $A'A$, $B'B$ or $C'C$, and $r_c$ depends on geodesics specified.

\begin{figure}[hpt]
\centering
\subfigure[]
{
\includegraphics[width=3in]{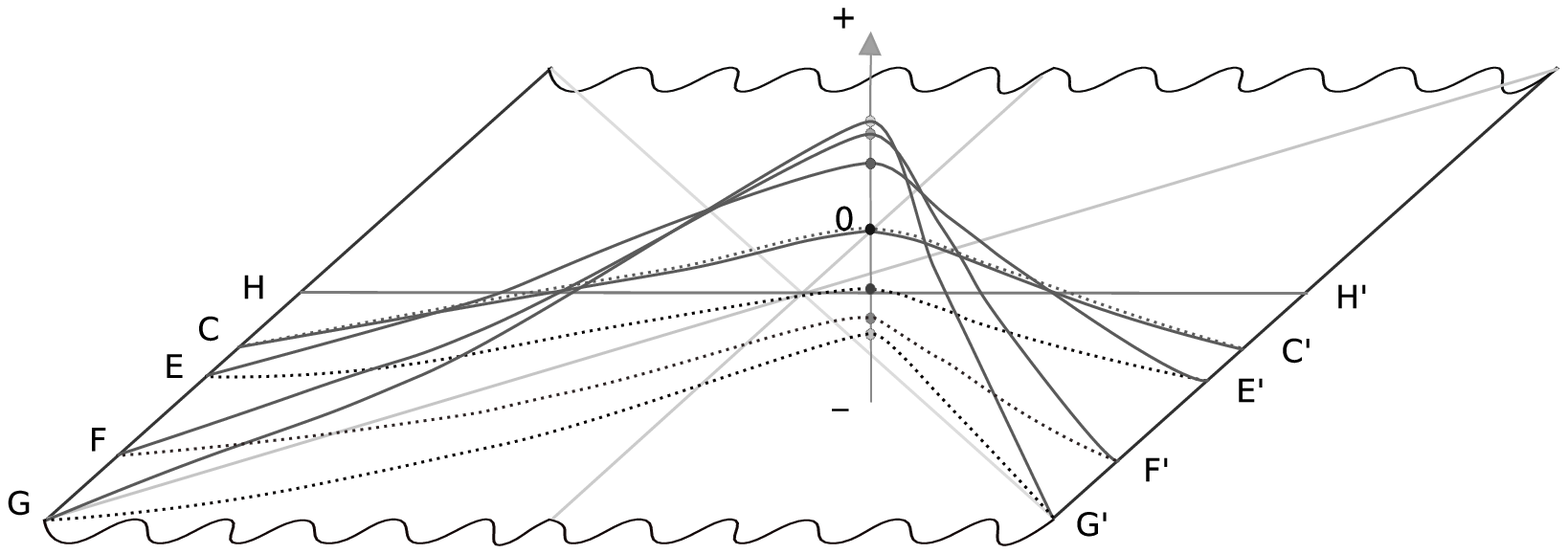}
}

\subfigure[]
{
\includegraphics[width=2in]{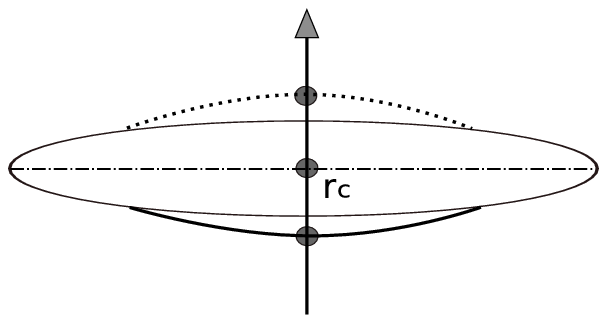}
}
\hspace{1.5cm}
\subfigure[]
{
\includegraphics[width=1in]{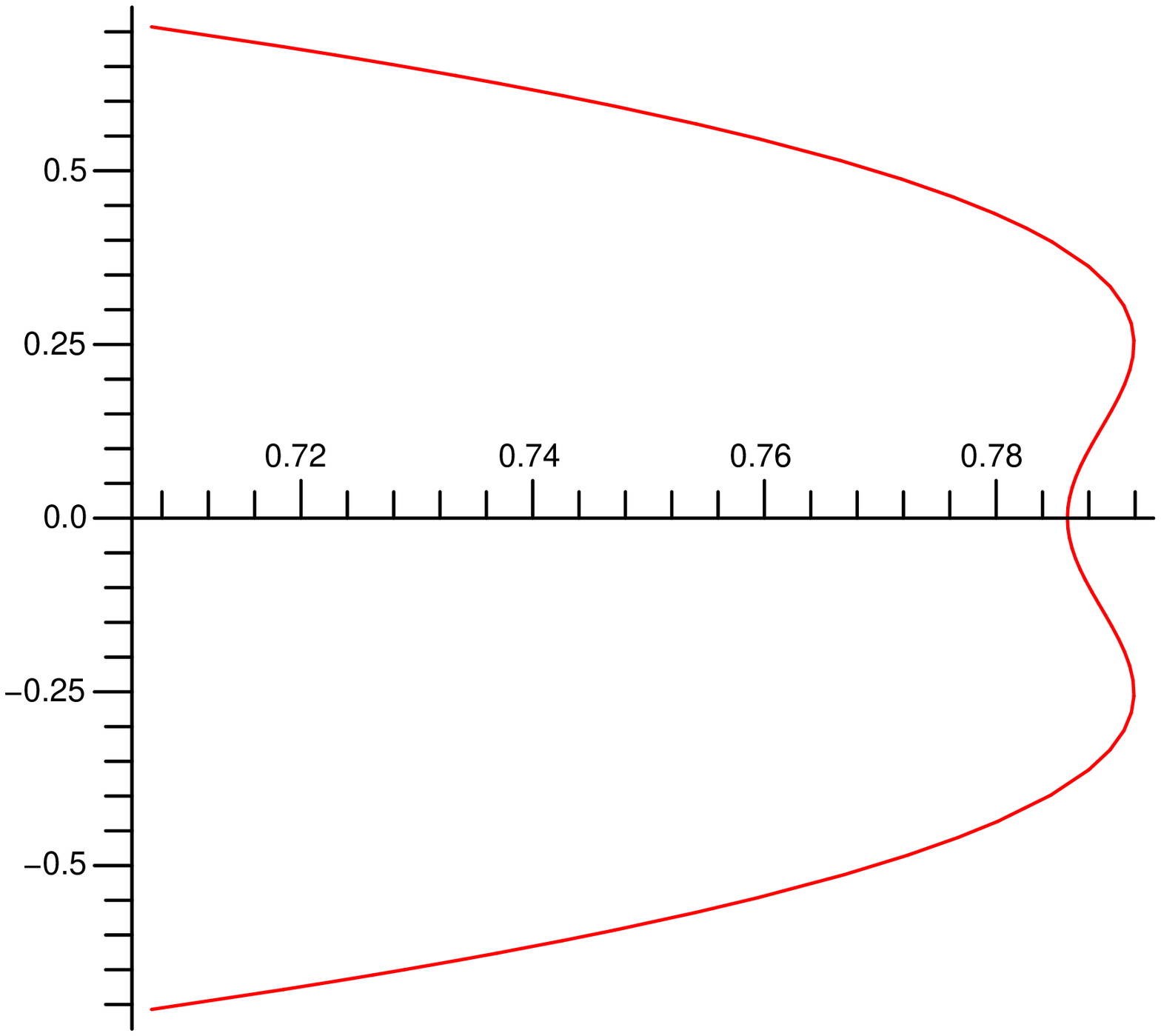}
}
\caption[]{The spacetime related to the cases of collision and after collision of  the two saddles. (a) The Lorentzian section. Solid eodesics  correspond to the left saddle, which is from the previous upper one. Dashed geodesics correspond to the right saddle, which is from the previous lower one. (b) The Euclidean section. After the collision of saddles, there are two geodesics dominating simultaneously. The solid geodesic corresponds to the left saddle and the dashed  geodesic corresponds to the right saddle. (c) After collision, the turning points pick up imaginary part of opposite sign. Real part of $r_c$ is in the horizontal direction and imaginary part of $r_c$ is in the vertical direction. As time increases, the two points start from the real axis, and split to upper and lower direction in a symmetrical way. This is the precise image of the vertical arrowed lines in (a) and (b).}
\label{lorengeo2}
\end{figure}
Before the collision of the saddles, the dominating geodesic is always the solid one. After the saddles collide, the saddles pick up real part frequencies at the same time and both saddles make an equal contribution in the final approximation. (See Fig. \ref{fit3}). Correspondingly, there appear two dominating geodesics. They both have Euclidean time difference with respect to the reference geodesic $G_0$. However, since their differences is of opposite sign, the summation of their Euclidean time difference with respect to $G_0$  remains $0$. For both of the geodesics, their turning points are no longer real valued. So we need to extend both the Euclidean section and the Lorentzian section in a third dimension to indicate the direction of the imaginary part of the turning points $r_c$ as shown in (a), (b) of Fig. \ref{lorengeo2}. The magnified version of the third dimension is shown in (c) of Fig. \ref{lorengeo2}.
When the saddles collide, it is indicated as $C'C$ in (a) of Fig. \ref{lorengeo2}. The two very close geodesics still lie inside the two-dimension Penrose diagram, with their real turning points inside the horizon. Let solid geodesics correspond to the left saddle, which is from the previously upper one and dashed geodesics correspond to the right saddle, which is from the previously lower one.  As time separation increases from $C'C$, to $E'E$, $F'F$ and then to $G'G$,  turning points of the solid geodesic now slowly rise above the Penrose diagram. Their turning points have almost constant real part and increasing imaginary part. This is shown as the upper half of (c) in Fig. \ref{lorengeo2}. 
At the same time, the turning points of the  dashed geodesic now slowly descends to underneath the Penrose diagram. They are  again with almost constant real part, but decreasing imaginary part. It is shown as the lower half of (c) in Fig. \ref{lorengeo2}.

The Euclidean sections of these geodesics are less simple to describe. For the pair of dominating geodesics, (except for $C'C$,) in (a) of Fig. \ref{lorengeo2}, there are two corresponding  geodesics in the Euclidean section as their projections. The two images have opposite time differences with respect to $-i\frac{\beta}{2}$ and with same real parts and opposite imaginary parts of turning points. They are indicated in (b) of Fig. \ref{lorengeo2}.  The solid geodesic is the projection of the solid geodesics from  (a) and the  dashed geodesic is the projection of dashed one from (a). The centre of the disk is decided by the real part of the turning points of various geodesics. 

In summary, for the time separation being Euclidean as $i\tau-i\beta/2$,  one saddle point appears. This implies that for the geodesic approximation, there is one dominant  geodesic for each $\tau$. This geodesic has purely Euclidean proper time $i\tau-i\beta/2$. As $\tau$ increase, the turning point of the geodesics varies from the horizon to the infinity. For the time separation being $t-i\beta/2$, there are always two saddle points appearing as solutions of the saddle point equation. However, they don't both dominate all the time. As $t$ increases from $0$ to the time when the two saddles collide, only the upper saddle dominates. The corresponding geodesic approximation tells us that the dominating geodesic starts from the one with time separation $-i\beta/2$ and turning point at the event horizon, i.e. $G_0$.    As $t$ increases, the turning point of the geodesic moves inside to the horizon. At the same time, the turning point of the other (not dominant) geodesic is moving from deep inside the singularity outwards. We will make more comment in the next paragraph on the non-dominating partner. By the time the saddles collide, the two geodesics almost coincide. After that, two geodesics contribute equally in the geodesic approximation. They always have the same real part of turning point and opposite sign  imaginary parts of the turning point. As $t$ increases,  real parts of their turning points stay inside the horizon not changing much, around $0.7$ to $0.8$ (see (c) of Fig. \ref{lorengeo2}), while imaginary parts increase in same amount in norm. In the Euclidean section, the two dominating geodesics have time difference $-i\beta/2+i\delta$ and $-i\beta/2-i\delta$, rather than before when there is only one dominating geodesic with time separation $-i\beta/2$. We have seen that some of the dominating geodesics depending on the time separation  have turning points inside the horizon. In this sense, we say that the boundary correlation function encodes locations inside the event horizon. On the other hand, in all the cases, the turning points never go infinitely closed to the singularity. This surely is not good enough for the purpose of resolving singularity. 

As we mentioned, when the time difference is $t-i\beta/2$ for $t$ quite small, there appears the lower saddle. It moves from $-i\infty$ upwards along the imaginary axis. Its steepest descent direction is always along the imaginary axis. Recall that at the same time the upper saddle always has steepest descent direction in the horizontal direction. This implies that when the steepest descent method is applied, the lower one is not picked up because of the contour of integration is along the real axis. We can define a function from  the original Fourier integration by replacing the contour of integration by the imaginary axis as an analogue of the new observable proposed in \cite{festuccia-2006-0604} for $AdS_5$ black hole. Despite the fact that the new integration actually blows up and assuming a regulization is available, we study it  as a regularized mathematical function.  If we apply the steepest descent method as before,  the lower saddle rather than the upper one will become  dominant. That is to say, if this new function has any physical meaning to anybody, wavepackets can move along the geodesic with turning point deep inside the horizon at the semi-classical limit.  The smaller the time separation $t$ is, the deeper the particle can go. As $t$ approaches zero, the turning point approaches infinitely close to the singularity. In this way, the newly defined function as a function on the boundary carries information from the singularity and may help in resolving the BTZ black hole singularity. At the same time, we should admit that its physical meaning is not clear.

\section{Conclusion}
The main objective of this paper is to see how the two-point function in the boundary CFT carries signatures from inside the horizon of the BTZ black hole in the context of the AdS/CFT correspondence. By using the key identification between frequencies and geodesics in the semi-classical limit in \cite{festuccia-2006-0604}, we deduce a geodesic approximation from the saddle point approximation.
As an application, we compute the Hartle-Hawking Green function by mode summation directly and obtain the boundary two-point function by the AdS/CFT correspondence. By taking the semi-classical limit, we find saddle points of the two-point function with respect to the time separation and thus relate the saddle points to specific dominating geodesics on the bulk. There are two interesting features which appear. The first occurs when the time separation is $t-i\beta/2$, for the Lorentzian time separation  $t$ not very small. There are two dominant geodesics with turning points  complex conjugate of each other and also with Euclidean time separations $-i\beta/2+i\delta$ and $-i\beta/2-i\delta$, respectively. This indicates the approximation of the two-point function is given by two complex geodesics together. The second interesting feature appears when the time separation is $t-i\beta/2$ for the Lorentzian time separation $t$ not very far from zero. If we  replace the original integration contour in the Fourier integration of the position space two-point function by the imaginary axis and regularize the new integration, then the saddle picked up with respect to the new contour is the one who can have turning point infinitely closed to the singularity. If there is good explanation about the newly defined function in the boundary theory, it may help to resolve the singularity.

\section*{Acknowledgments}
I want to thank my supervisor Mukund Rangamani for constant help through the process of writing this paper. I indebted him for firstly showing me the main paper \cite{festuccia-2006-0604}, on which this work is based, secondly providing the method and computation of the momentum space correlation function by modes sum, where the constant $C(\omega, J)$ is  crucial for the later analysis on saddle points, and thirdly for useful discussions and comments in the layout this paper. I want to thank Derek Harland for lots of stimulating discussions and reading the manuscript. 

\section*{Appendix}
W will fix constants $C_1$ and $C_2$ so that the boundary conditions (\ref{bcs}) are satisfied. We will get an approximate form of the solution around boundary first and fix the constraint between $C_1$ and $C_2$ by imposing the boundary condition (\ref{bcs}).
As $r\longrightarrow \infty$, $z\sim\frac{1}{r}\longrightarrow 0$, the approximated solution is
$$X_{bdry}(z)
=(-1)^{-\frac{i\omega}{2}}z^{-\frac{1}{2}}z^{i\omega}(C_1 z^{iJ}G(z)+C_2 z^{-iJ}H(z)),
$$
where
$$
G(z)=F(\frac{-iJ-i\omega +1-\nu}{2}, \frac{-iJ-i\omega+1+\nu}{2}; 1-iJ, \frac{1}{z^2}),
$$
$$
H(z)=F(\frac{iJ-i\omega+1-\nu}{2}, \frac{iJ-i\omega+1+\nu}{2}; 1+iJ, \frac{1}{z^2}).
$$
From general identities of hypergeometric function \cite{Abramowitz}, we can further approximate the solution by
\begin{eqnarray}
\label{a5}
X_{bdry}(z)
&=&(-1)^{-\frac{1}{2}}((C_1 (-1)^{-\frac{-iJ-\nu}{2}}\Gamma_1+C_2 (-1)^{-\frac{iJ-\nu}{2}}\Gamma_2) z^{\frac{1}{2}-\nu}\\
&&+(C_1(-1)^{-\frac{-iJ+\nu}{2}}\Gamma_3+C_2(-1)^{-\frac{iJ+\nu}{2}}\Gamma_4) z^{\frac{1}{2}+\nu}),
\end{eqnarray}
where 
$$\Gamma_1=\frac{\Gamma(1-iJ)\Gamma(\nu)}{\Gamma(\frac{1}{2}(1-i\omega-iJ+\nu))\Gamma(\frac{1}{2}(1+i\omega-iJ+\nu))},$$
$$\Gamma_2=\frac{\Gamma(1+iJ)\Gamma(\nu)}{\Gamma(\frac{1}{2}(1-i\omega+iJ+\nu))\Gamma(\frac{1}{2}(1+i\omega+iJ+\nu))},$$
$$\Gamma_3=\frac{\Gamma(1-iJ)\Gamma(-\nu)}{\Gamma(\frac{1}{2}(1-i\omega-iJ-\nu))\Gamma(\frac{1}{2}(1+i\omega-iJ-\nu))},$$
$$\Gamma_4=\frac{\Gamma(1+iJ)\Gamma(-\nu)}{\Gamma(\frac{1}{2}(1-i\omega+iJ-\nu))\Gamma(\frac{1}{2}(1+i\omega+iJ-\nu))}.$$

Imposing the boundary condition  by requiring $z^{\frac{1}{2}-\nu}$ mode of (\ref{a5}) vanish, we get the constraint between the constant $C_1$ and $C_2$ as
\begin{equation}
\label{a6}
C_2=-(-1)^{iJ}\frac{\Gamma_1}{\Gamma_2}C_1.
\end{equation}
Thus, (\ref{a5}) becomes
\begin{equation}
\label{7}
X_{bdry}(z)=C(\omega,J)z^{\frac{1}{2}+\nu},
\end{equation}
where
\begin{eqnarray*}
C(\omega,J)&:=&C_1(-1)^{\frac{iJ-\nu-1}{2}}
(\Gamma_3-\frac{\Gamma_1}{\Gamma_2}\Gamma_4)\\
&=&C_1(-1)^{\frac{iJ-\nu}{2}}
\frac{1}{\pi^2}\sin(\pi\nu)\sinh(\pi J)\\
&&\Gamma(-\nu)\Gamma\left(\frac{1-i\omega+iJ+\nu}{2}\right)\Gamma(1-iJ)\Gamma\left(\frac{1+i\omega+iJ+\nu}{2}\right). 
\end{eqnarray*}
Next, we want to get the approximation of (\ref{RKG}) around the horizon and impose boundary conditions to fix constants $C_1$ and $C_2$ completely.

Again by general properties of hypergeometric functions, the approximation of (\ref{RKG}) around the horizon is 
\begin{eqnarray*}
X_{hor}(r)
&=&r^{\frac{1}{2}}((C_1\Gamma_5+C_2\Gamma_6)(1-r^2)^{-\frac{i\omega}{2}}+(C_1\Gamma_7+C_2\Gamma_8)(1-r^2)^{\frac{i\omega}{2}}),
\end{eqnarray*}
where
$$\Gamma_5=\frac{\Gamma(1-iJ)\Gamma(i\omega)}{\Gamma(\frac{1}{2}(1+i\omega-iJ+\nu))\Gamma(\frac{1}{2}(1+i\omega-iJ-\nu))},$$
$$\Gamma_6=\frac{\Gamma(1+iJ)\Gamma(i\omega)}{\Gamma(\frac{1}{2}(1+i\omega+iJ+\nu))\Gamma(\frac{1}{2}(1+i\omega+iJ-\nu))},$$
$$\Gamma_7=\frac{\Gamma(1-iJ)\Gamma(-i\omega)}{\Gamma(\frac{1}{2}(1-i\omega-iJ-\nu))\Gamma(\frac{1}{2}(1-i\omega-iJ+\nu))},$$
$$\Gamma_8=\frac{\Gamma(1+iJ)\Gamma(-i\omega)}{\Gamma(\frac{1}{2}(1-i\omega+iJ-\nu))\Gamma(\frac{1}{2}(1-i\omega+iJ+\nu))}.$$
In the $z$ coordinates, 
\begin{eqnarray*}
X_{hor}(z)
&=&\exp(-\frac{\pi\omega}{2})[(C_1\Gamma_5+C_2\Gamma_6)2^{-i\omega}\exp(-i\omega z)\\
&&+(C_1\Gamma_7+C_2\Gamma_8)\exp(\pi\omega)2^{i\omega}\exp(i\omega z)].
\end{eqnarray*}
By the constraint (\ref{a6}) from boundary, 
\begin{eqnarray*}
X_{hor}(z)&=&\exp(-\frac{\pi\omega}{2})[(C_1\Gamma_5-(-1)^{iJ}\frac{\Gamma_1}{\Gamma_2}C_1\Gamma_6)2^{-i\omega}\exp(-i\omega z)\\
&&+(C_1\Gamma_7-(-1)^{iJ}\frac{\Gamma_1}{\Gamma_2}C_1\Gamma_8)\exp(\pi\omega)2^{i\omega}\exp(i\omega z)]\\
&=&C_1 \exp(-\frac{\pi\omega}{2})(2^{-i\omega}P \exp(-i\omega z)+ 2^{i\omega}Q \exp(i\omega z)),
\end{eqnarray*}
where 
$$
P=\frac{(\exp(J\pi)-\exp(-J\pi))\Gamma(1-iJ)\Gamma(i\omega)}{(\exp(J\pi)+\exp(\pi(\omega+i\nu)))\Gamma(\frac{1}{2}(1+i\omega-iJ-\nu))\Gamma(\frac{1}{2}(1+i\omega-iJ+\nu))},
$$
$$
Q=\frac{\exp(-\pi\omega)(\exp(J\pi)-\exp(-J\pi))\Gamma(1-iJ)\Gamma(-i\omega)}{(\exp(J\pi)+\exp(\pi(-\omega+i\nu)))\Gamma(\frac{1}{2}(1-i\omega-iJ-\nu))\Gamma(\frac{1}{2}(1-i\omega-iJ+\nu))}.
$$
Numerically, it turns out that  $|P|=|Q|$.
 The boundary condition at the horizon implies that $C_1$ should be chosen as $C_1=\exp(\frac{\pi\omega}{2})(PQ)^{-\frac{1}{2}}$.
Indeed,
\begin{eqnarray*}
X_{hor}(z)&=&\frac{\exp(\frac{\pi\omega}{2})}{\sqrt{PQ}} \exp(-\frac{\pi\omega}{2})(2^{-i\omega}P \exp(-i\omega z)+ 2^{i\omega}Q \exp(i\omega z))\\
&=&2^{-i\omega}\sqrt{\frac{P}{Q}} \exp(-i\omega z)+ 2^{i\omega}\sqrt{\frac{Q}{P}} \exp(i\omega z).
\end{eqnarray*}
If we define $\theta$ by $\exp(i\theta)=\sqrt{\frac{Q}{P}}$, then $|P|=|Q|$ implies that $\theta$ is real. Let $\delta=\omega \ln 2+\theta$, then the solution is $X_{hor}(z)=\exp(i\omega z)\exp(i\delta)+\exp(-i\omega z) \exp(-i\delta)$, and hence satisfies the boundary condition. After $C_1$ is fixed, $C(\omega, J)$ is fixed and its square is given as,
\begin{eqnarray*}
C(\omega, J)^2&=&\frac{1}{2\pi^3}\exp(\pi\omega)(\exp(2\pi\omega)-1)\omega\sin^2(\pi\nu)\Gamma(-\nu)^2\\
&&\Gamma(\frac{1}{2}(i\omega+iJ+\nu+1))\Gamma(\frac{1}{2}(-i\omega+iJ+\nu+1))\nonumber\\
&&\Gamma(\frac{1}{2}(i\omega-iJ+\nu+1))\Gamma(\frac{1}{2}(-i\omega-iJ+\nu+1)).\nonumber
\end{eqnarray*}

\bibliographystyle{unsrt}
\bibliography{BTZsing}

\end{document}